%% file: MM-Auger.tex
\PassOptionsToPackage{sort&compress}{natbib}
\documentclass[final]{frontiersFPHY} 

\usepackage{url,hyperref,lineno,microtype,subcaption}
\usepackage[onehalfspacing]{setspace}
\newenvironment{description}{}{}
\usepackage{enumitem}
\usepackage{float}

\def\keyFont{\fontsize{8}{11}\helveticabold}
\def\firstAuthorLast{A.~Aab {et~al.}} 
\def\Authors{The Pierre Auger Collaboration\,$^{1}$}
\def\Address{$^{1}$Pierre Auger Observatory, Av.\ San Mart\'{i}n Norte 304, 5613 Malarg\"ue, Argentina\newline
\textnormal{Author names and affiliations appear at the end of this paper.}}
\def\corrAuthor{~}
\def\corrEmail{auger\_spokespersons@fnal.gov}

\begin{document}
\onecolumn
\firstpage{1}

\title[Multi-Messenger Physics with the Pierre Auger Observatory]{Multi-Messenger Physics with the Pierre Auger Observatory} 

\author[\firstAuthorLast]{\Authors} 
\address{} 
\correspondance{} 

\extraAuth{}

\maketitle

\begin{abstract}
An overview of the multi-messenger capabilities of the Pierre Auger Observatory is presented. The techniques and performance of searching for Ultra-High Energy neutrinos, photons and neutrons are described. Some of the most relevant results are reviewed, such as stringent upper bounds that were placed to a flux of diffuse cosmogenic neutrinos and photons, bounds placed on neutrinos emitted from compact binary mergers that were detected by LIGO and Virgo during their first and second observing runs, as well as searches for high energy photons and neutrons from the Galactic center that constrain the properties of the putative Galactic PeVatron, observed by the H.E.S.S.\ collaboration. The observation of directional correlations between ultra-high energy cosmic rays and either high energy astrophysical neutrinos or specific source populations, weighted by their electromagnetic radiation, are also discussed. They constitute additional multi-messenger approaches aimed at identifying the sources of high energy cosmic rays.
\end{abstract}

{\keyFont{\section*{Keywords:} UHECR, high energy neutrinos, high energy photons, high energy neutrons, multi-messenger astrophysics, compact binary mergers}}

\input{Intro-Auger}
\input{Neutrinos}
\input{Photons}
\input{Neutrons}
\input{Neutrino-UHECR-corr}
\input{Conclusions}


\section*{Conflict of Interest Statement}

The authors declare that the research was conducted in the absence of any commercial or financial relationships that could be construed as a potential conflict of interest.

\section*{Author Contributions}
All authors contributed to the original material and writing of the manuscript. KHK and EZ coordinated this review.

\input{acknowledgments}

\subsection*{Full author list and affiliations}
\footnotesize
\input{latex_authorlist_authors}

\input{latex_authorlist_institutions}
\normalsize



\bibliographystyle{frontiersinHLTH_FPHY} 
\bibliography{references}


\end{document}

%% file: Intro-Auger.tex
\section{Introduction}


With the discovery of neutrinos from SN1987A~\citep{Hirata:1987hu,Bionta:1987qt} arriving four hours before the 
light detected by conventional telescopes it became clear that there was a
lot to learn from examining any type of particles and radiation coming from
astrophysical objects, and that neutrino detectors could give early alerts that
would facilitate the observation of the evolution of such transients from the
earliest stages. Besides the different
wavelengths of traditional astronomy, neutrinos, 
cosmic rays, very high energy gamma rays, and gravitational waves provide complementary information to study the most energetic objects of the Universe. While the SN1987A event might 
be said to mark the onset of multi-messenger 
astronomy, the term was only introduced in the late 1990s
(see, e.g.\ \citep{Barwick:1998xq,Halzen:2003mw}), and the final boost to the field took place quite recently with the emergence of both neutrino~\citep{Aartsen:2014gkd} and gravitational-wave astronomy~\citep{Abbott:2016blz}. 
Indeed the discovery of gravitational waves from the merging of a neutron 
star binary system by LIGO and Virgo \citep{TheLIGOScientific:2017qsa} triggered a spectacular series of observations in the full electromagnetic spectrum from radio to the very high energy gamma 
rays \citep{GBM:2017lvd} and searches for  neutrino fluxes with ANTARES,
IceCube, 
and the Pierre Auger Observatory \citep{ANTARES:2017bia}. The combined effort marks an unprecedented leap forward in astrophysics revealing many aspects of the Gamma-Ray Burst (GRB) induced by the merger and its subsequent kilonova. More recently an energetic neutrino candidate was  detected in IceCube in coincidence with the powerful blazar TXS-0506+056 during a flare in the very-high-energy gamma-ray band, incompatible with a chance correlation at the $3\sigma$-level~  \citep{IceCube:2018dnn}. 
After scanning archival data, possible evidence for enhanced neutrino emission (quantified at $3.5\sigma$-level) was also found from this direction,  during an independent 5-month period between September 2014 and March 2015~\citep{IceCube:2018cha}. It is now quite clear that there is much potential in the combined analysis of data from different experiments and multi-messenger astronomy has excellent 
prospects of making many more significant contributions in the near future. 

The Pierre Auger Observatory is the largest and most precise detector of ultra-high energy air showers, such as those regularly induced by cosmic rays. 
While cosmic rays are expected to have curved trajectories in their path to Earth and thus lose time correlation with light emission, at the highest energies they could still keep directional information on the as-yet unidentified sources where they are produced. The study of the arrival directions of these particles has revealed deviations from isotropy \citep{Aab:2017tyv,Aab:2018mmi} that, combined with other particle searches,  could be of relevance for multi-messenger astronomy. Indeed spatial correlations have been searched for between the arrival directions of the highest energy cosmic rays and classes of objects that have been proposed as sources in the very high-energy regime such as Active Galactic Nuclei (AGN) and Starburst Galaxies (SBG) \citep{Aab:2018chp}. Notably, it has also been shown that it is possible to detect neutral particles of sufficiently high energy with the Pierre Auger Observatory. By looking at the depth development of the showers, it is relatively easy to identify neutrinos which interact in the lower layers of the atmosphere. Photons can also be discerned  from the background of cosmic rays because the produced showers have a reduced  number of muons and they  develop deeper than cosmic rays in the atmosphere. Finally, there  is no known possibility to separate neutron-induced showers from the charged cosmic  rays on the basis of the shower development but, since neutrons are directional, it is in principle possible to identify sources of nearby neutrons by looking at an excess from given directions or to exploit potential time and  directional correlations. This procedure could also be applied to any type of neutral particles that induce a shower in the atmosphere such as photons. 

In this article, we review the capability of the Observatory to search for
signals of such particles and discuss the contributions that have been made.

\section{The Pierre Auger Observatory}

The Pierre Auger Observatory is designed to detect extensive air showers produced by primary cosmic rays above 0.1\,EeV. It is located near the city of Malargüe, Argentina, at a latitude of $35.2^\circ$S, a longitude of $69.2^\circ$W and at an approximate altitude of 1400\,m above sea level, or equivalently an atmospheric depth $X_{\mathrm{ground}} = 880$\,g\,cm$^{-2}$. 
It comprises a surface detector (SD) array of 1660 water-Cherenkov stations deployed over a triangular grid of 1.5 km spacing and a system of 27 telescopes grouped in four sites forming the fluorescence detector (FD). The telescopes are erected at the periphery of the Observatory to observe the atmosphere over the full area of 3000\,km$^2$ covered by the SD array. The SD stations sample the density of the secondary particles of the air shower at the ground and are sensitive to the electromagnetic, muonic and hadronic components. The FD observes the longitudinal development of the air shower by detecting the fluorescence and Cherenkov light emitted during the passage of the secondary particles of the shower in the atmosphere. Unlike the SD, the fluorescence telescopes are operated only during clear and moonless nights, for an average duty cycle of about 14\,\%. The hybrid feature of the Auger Observatory combining these two well-established techniques has proved to be extremely rewarding in making it the most precise instrument to reconstruct the energy, mass, and direction of Ultra-High Energy Cosmic Rays (UHECRs). Details about the different detector components and their performances can be found in \citep{ThePierreAuger:2015rma,Abraham:2009pm}.

Besides its general hybrid capabilities,  a feature that makes the Pierre
Auger Observatory a very versatile and powerful multi-messenger observatory is
the specific design of the SD stations. They are constructed as cylinders of
3.6 m diameter and 1.2 m height filled with 12 tonnes of purified water,
each. Charged particles entering a station induce emission of Cherenkov light which is reflected at the walls by a diffusive Tyvek liner, and collected by three 9-inch photomultiplier tubes (PMT) at the top surface and in optical contact with the water. The PMT signals are sampled by flash analog-to-digital converters (FADC) with a time resolution of 25\,ns~\citep{Abraham:2009pm}\footnote{The recorded sequence of signals every 25 ns is referred to as the signal trace.}. This provides good discrimination of electrons and muons entering the detector station from the top, a feature which is important to identify muon-poor air showers induced by photons, not only with the FD but also with the SD. Moreover, the detector stations also provide a large cross section for inclined and up-going air showers, a feature that is of key-importance for the detection of neutrino-induced air showers. Both of these aspects will be discussed in more detail below.

The modular structure of the surface detector array and fluorescence
telescopes allowed the data taking to start in 2004 in a partial
configuration. In 2008 the detector was completed and since then 
data have been taken, practically without interruption.
Once installed, the local stations work practically without external 
intervention. 
There are two types of trigger conditions: a local trigger at the level of an individual station (second order or T2 trigger), and a global trigger (third order or T3 trigger) to register events.  The T2 trigger condition is satisfied when either the signal exceeds the equivalent of 3.2 Vertical Equivalent Muons (VEM) in at least one time bin of the signal trace in the FADC -- the so-called ``threshold trigger'' -- or when it exceeds a much lower threshold (0.2 VEM) in at least 13 bins within a 3 $\mu$s time window (i.e., 120 bins) -- the so-called “Time-over-threshold (ToT) trigger.”  The ToT condition was designed to trigger on signals broad in time, characteristic of the electromagnetic component dominant over the first 1000~g cm$^{-2}$ of the extensive air shower, and it is crucial for neutrino identification as explained below. The data acquisition system receives the local T2 triggers and builds a global T3 trigger requiring a relatively compact configuration of at least three local stations compatible in time, each satisfying the ToT trigger, or four triggered stations within a time window slightly larger than light travel time with any type of T2 trigger \citep{ThePierreAuger:2015rma}. With the completed array, the global T3 trigger rate is about two events per minute, one third being showers with energies above $3\times 10^{17}$\,eV. 

%% file: Neutrinos.tex
\section{Neutrinos}


Neutrinos can travel in straight paths from the confines of the Universe, are capable of going through large matter depths without absorption, and are excellent messengers from extragalactic sources, where protons or nuclei are thought to be accelerated to high energies. Neutrinos are also produced from cosmic-ray interactions with the cosmic microwave background. Besides pointing to  the position of their sources, they can also provide valuable information concerning hadronic acceleration processes, composition, the local environment at the sources, and their evolution with redshift~\citep{Gaisser:1994yf}. 

The idea of using inclined showers to search for neutrino interactions is old~\citep{Berezinsky:1975zz}, and the large potential of the Pierre Auger Observatory for neutrino detection was already recognized in its design stages~\citep{Capelle:1998zz}. 
Inclined showers induced by cosmic rays were observed in Haverah
Park~\citep{Hillas:1969zza}, in the early days of extensive air shower
arrays. It was at the beginning of the 2000s when the muon patterns of the 
showers at ground level were sufficiently understood to allow shower
reconstruction~\citep{Ave:2000xs} (see also~\citep{Aab:2014gua}). These
showers traverse large atmospheric depths, and their electromagnetic component
gets almost completely absorbed before the shower particles reach ground
level. As a result, for zenith angles exceeding $\sim 60^\circ$,
the shower front reaching the SD detector stations consists mainly of muons that typically have energies between 20 and 200 GeV and travel tens of km without decaying. These muons are little affected by interactions, except for continuous energy loss and deflections in the Earth's magnetic field~\citep{Ave:2000xs}. As their radius of curvature is a few thousands km, they do not deviate much from their initial trajectories, keeping the timing of the shower front sharp. 

As opposed to cosmic rays, inclined neutrinos can interact deep in the
atmosphere because their interaction length exceeds the matter depth of
the atmosphere for any zenith angle, $\theta$. The resulting ``young'' showers 
are rich in electrons and photons at ground level. These showers induce signals 
typically spread over much wider time intervals than cosmic rays with the same 
$\theta$ because of multiple scattering of the electrons and
photons in the shower front. The broader digitized signal traces recorded at 
the particle detectors of the SD array resemble those 
produced by vertical air showers and it is thus relatively easy to single them
out from the sharply arriving fronts produced by inclined protons or nuclei. 
Tau neutrinos interact in the Earth just below the surface and produce tau
leptons that escape into the atmosphere, decaying in flight and producing 
up-going air showers~\citep{Fargion:2000iz,LetessierSelvon:2000kk}. For 
neutrino energies exceeding 100 PeV, the combined conversion and exit 
probability is maximal for nearly horizontal (``Earth-skimming'') directions 
that exit with an elevation angle between $1^\circ$ and $2^\circ$ above the 
horizontal direction~\citep{Zas:2005zz}. These showers develop close and
almost parallel to the ground. As the shower particles, particularly electrons
and photons, spread laterally they can reach the SD stations, inducing 
characteristic signals which are also easily identified because of their 
broader time spread. 

\subsection{Selection and identification}
\label{s:NuIdentification}

The event selection, the neutrino identification, and the exposure
calculations are performed separately for Earth-skimming (ES) and down-going 
(DG) showers~\citep{Aab:2015kma}. To select ES showers, all triggered events 
involving at least three stations are considered. The ellipsoidal signal 
pattern at the ground is required to have a large eccentricity (characterized 
by a ratio of major and minor axes greater than $5$), and the direction of the
major axis indicates the azimuthal angle of the event. The apparent speed of 
the signal at the ground, measured for each pair of stations and averaged over 
all of the pairs, must be in the interval $[0.967,1.034]c$~\footnote{Actual 
values used are [0.29, 0.31] m~ns$^{-1}$}, and the r.m.s.\ spread must be less 
than $0.267c$~\footnote{0.08 m~ns$^{-1}$}. For the DG
selection~\citep{Abreu:2011zze} a minimum of four triggered stations is
required to reduce noise from random triggers, and two zenith angle groups are 
made:``Low'' (DGL) and ``High'' (DGH) respectively for $60^\circ<\theta<75^\circ$ 
and $75^\circ<\theta<90^\circ$. 
The selection of DGL showers is just made requiring the reconstructed zenith angle 
to be in the $58.5^\circ$-$76.5^\circ$ range allowing for uncertainties. 
It is obtained fitting a plane to the positions of the triggered stations and the start time of the signals. For the selection of DGH showers, the 
reconstructed $\theta$ must exceed $75^\circ$, the apparent average speed of the signal at the ground must be smaller than 
$1.044c$~\footnote{0.313 m~ns$^{-1}$}, its spread must be smaller than 
$0.08c$, and the eccentricity must exceed a value of 3. 

The ES (DG) neutrino identification has been optimized using extensive shower
simulations for different energies, arrival directions, decay (interaction)
depths, and impact parameters, with standard software such as 
AIRES~\cite{Aires:1999} and CORSIKA~\cite{Heck:2000} and including the SD 
response of the individual detector stations using specific 
software~\citep{Argiro:2007qg,ThePierreAuger:2015rma} based on 
GEANT4~\cite{Agostinelli:2002hh}. A single variable has been chosen for the
identification procedure in both searches. Its distribution for neutrino
simulations is compared to that of regular showers obtained from a small
sample of the data assumed to be exclusively cosmic rays. To select neutrino
candidates, a cut is made on this variable at a value that  would just yield
one cosmic-ray in 50 years according to the extrapolated background
distribution. For instance, in the ES case, this variable is the average area
over peak ($AoP$) value of the signal traces of each event. The area is the
integrated charge of the signal trace and the peak is its maximum value, both 
normalized to one for vertical throughgoing muons used for on-line calibration. 
The cut obtained is $AoP>1.83$, ensuring sufficient time spread in the signal
to reject cosmic-ray showers~\citep{Aab:2015kma}. Events with just three
stations are further required to have $AoP$ exceeding 1.4. 
For DG showers, a Fisher discriminant approach combines several variables 
based on the $AoP$ of a few selected stations. For the DGH case, the
multivariate Fisher method~\cite{Fisher:1936et} uses nine variables obtained
with the $AoP$ values of the four earliest stations and a tenth variable
accounting for the asymmetry in the time distribution of the earliest and
latest stations of the event. For the DGL case, the Fisher variable combines
five or six variables obtained from the $AoP$ of the four or five stations
closest to the shower core and their product. For DGL showers, $75\%$ of the
triggers are also required to be of special ``ToT''
kind~\citep{Aab:2015kma}. For optimization purposes of DG neutrinos, further
subdivisions have been made using the number of triggered stations for the DGH
set and the reconstructed $\theta$ for the DGL set. The condition for a
neutrino candidate is slightly different in each subset. 

\begin{figure}[t]
\centering
\includegraphics[width=.49\linewidth]{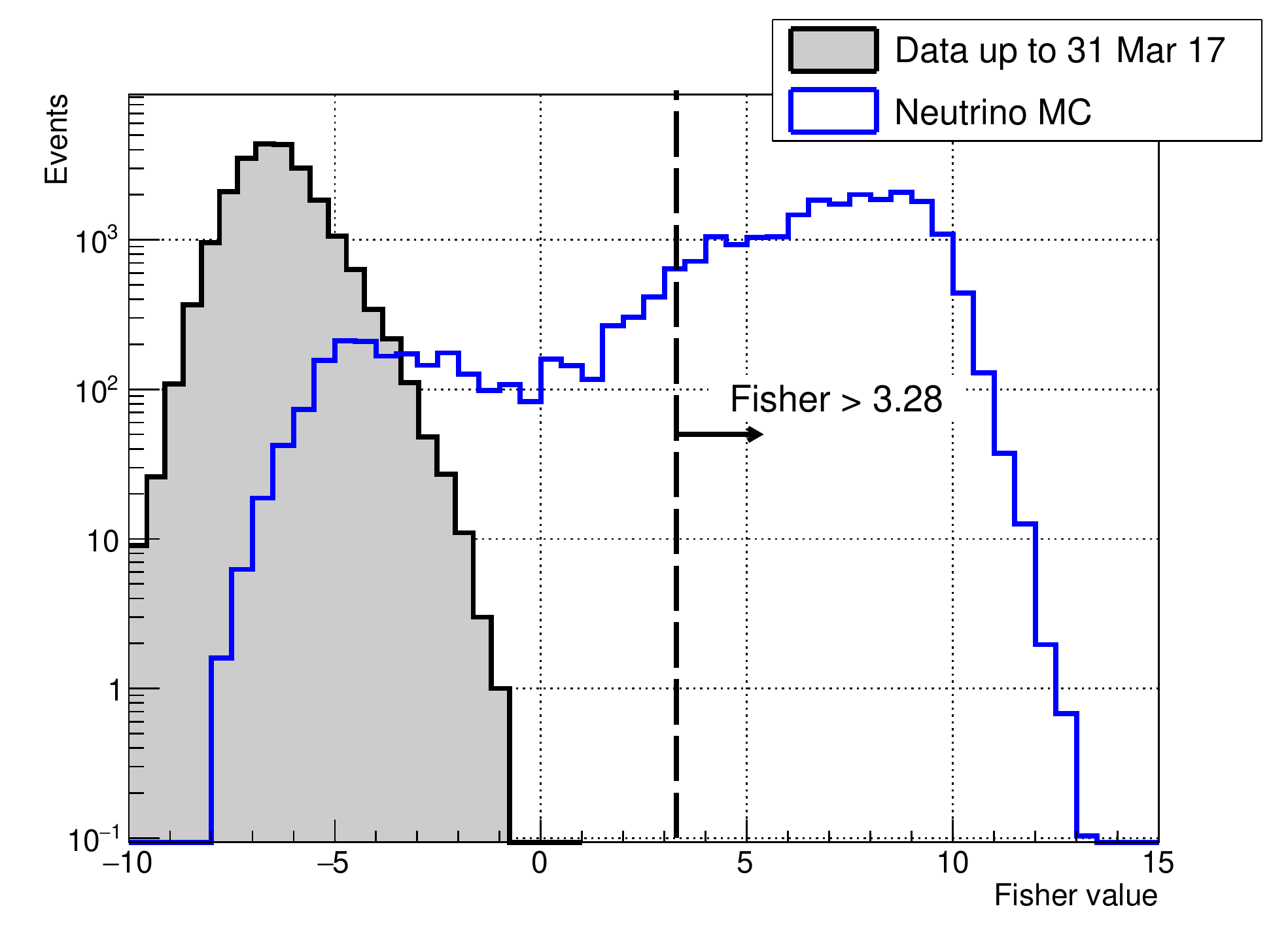}
\includegraphics[width=.49\linewidth]{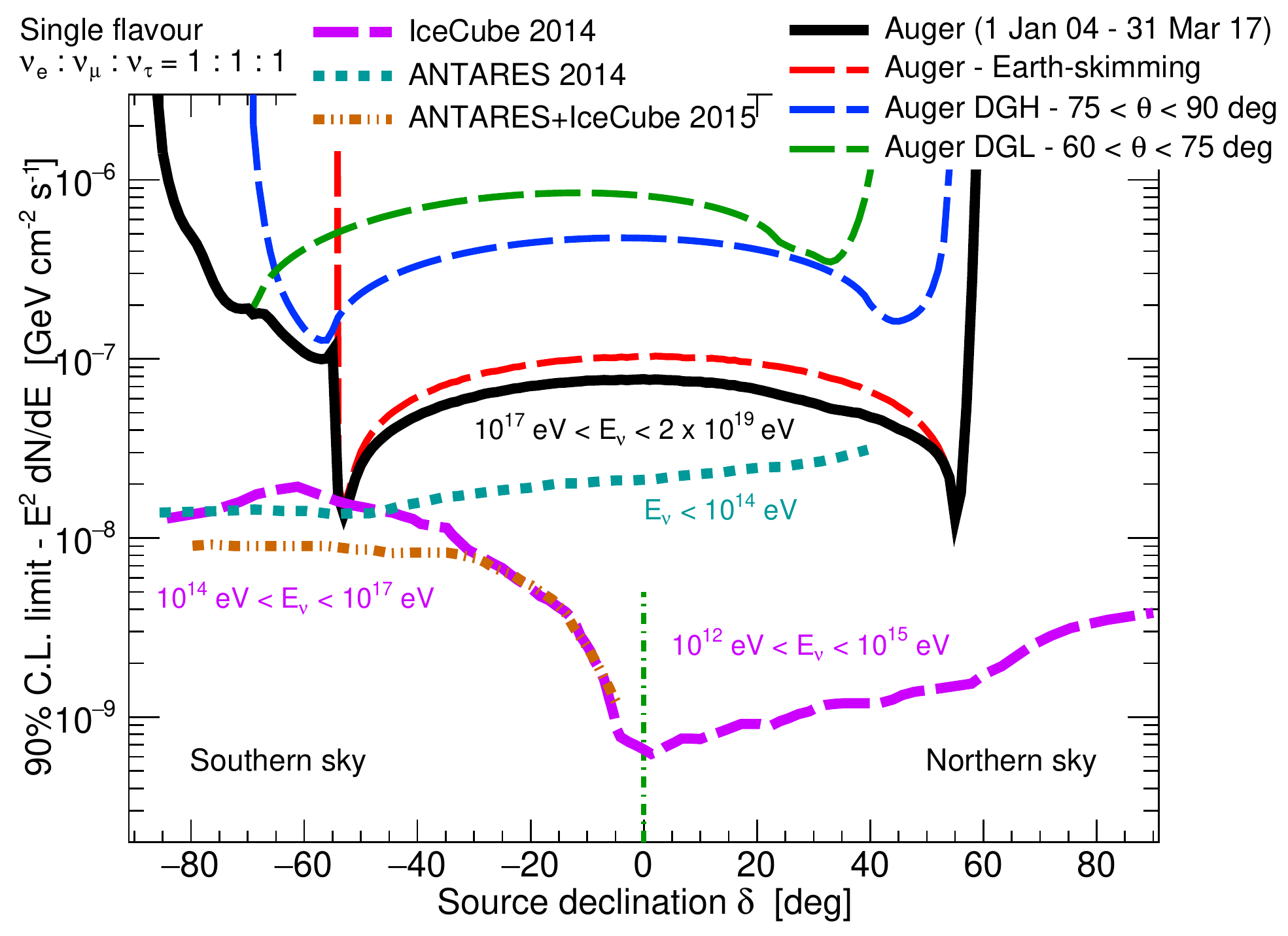}
\caption{Left: Distribution of the Fisher variable for the DGH search (see text) for simulations compared to a small fraction of the data and assumed to be due to cosmic rays. The cut is made at a value of the Fisher variable of 3.28. Right: Limits to the point source fluxes as a function of equatorial declination obtained from the non-observation of ES and DGH neutrino candidates up to March 31$^{\rm st}$ 2017 (from~\citep{Zas:2017xdj}).}
\label{fig:neutrino-search}
\end{figure} 

\subsection{Efficiency and Aperture}
\label{s:Aperture}

The effective aperture of the detector (aperture from now on) is defined so
that when it is multiplied by the neutrino spectral flux, it gives the energy 
distribution of the detection rate. For point sources in the DG case, 
this aperture is simply calculated integrating the interaction and detection 
probabilities over the SD 
area. Three separate apertures are calculated according to
interaction type and neutrino flavor: 1) neutral current for all flavors
and charged current muon neutrinos, where the shower is produced by the
nuclear debris (typically carrying about $20\%$ of the neutrino energy);
2) charged current for electron neutrinos where all the energy is converted to
the air shower ($80\%$ induced by the electron and the rest by nuclear
fragments); 3) charged current for tau neutrinos where the shower can be
produced either by the nuclear debris ($20\%$ of the energy) or, deeper in 
the atmosphere, by the tau decay ($80\%$ of the energy). 
The three apertures can be added under the
assumption of equal fluxes for each neutrino flavor, as expected because 
of neutrino flavor oscillations in the characteristic long baselines provided 
by extragalactic distances to plausible
sources~\citep{Athar:2000yw,Learned:1994wg}. 

On the other hand, only tau
neutrinos can be efficiently detected in the ES mode. When muon neutrinos 
convert to high energy muons that exit the Earth, practically
all of them also exit the atmosphere without decaying because of their
longer lifetime and when electron neutrinos convert to electrons the shower 
develops rapidly in the Earth before exiting. 
In the ES case, the aperture is calculated as an integral over the Earth's 
area (transverse to the flux direction), weighted by the probability of a 
tau neutrino interacting and producing a tau which subsequently exits the Earth,
decays in the atmosphere and induces a shower that triggers the SD and
is effectively selected and identified. This is done in two steps. 
The differential probability of a tau lepton of given
energy exiting the Earth has been calculated as a function of $\theta$ using
simulations of tau neutrino interactions in rock that include
regeneration~\cite{Abreu:2013zbq}. The tau-exit probability must be integrated 
over decay distance weighted by the survival probability, the decay 
probability per unit distance and the probability of detection with the SD. 
The detection probability for both DG and ES showers includes 
trigger, selection and identification and it is also obtained from the
simulations.

%
%
The exit and detection probabilities in the aperture calculation depend
strongly on $\theta$. A given source located at right ascension $\alpha$, 
and declination $\delta$ in equatorial coordinates, is observed with a zenith
angle that oscillates with time according to: 
\begin{equation}
\cos \theta (t) = \sin \lambda \sin \delta + \cos \lambda \cos \delta \sin (2\pi t/T  - \alpha).
\end{equation}
Here $\lambda=-35.2^\circ$ is the latitude of the detector, $t$ is the
sidereal time, and $T$ the duration of a sidereal day. 
Naturally, the aperture is strongly dependent on source position and 
time. At any given instant the field of view of the Observatory for each of
the ES, DGH and DGL channels is limited to the bands corresponding to the
zenith angle range of the channel, as displayed in 
Figs.~\ref{fig:fov_GW150914} (left) and \ref{fig:fov_GW170817} for specific
examples discussed below. 

%

Typically, the search for neutrinos from point sources is performed on a
pre-selected time interval chosen to match plausible emission times 
according to theoretical models describing the mechanisms acting within 
the sources. The expected event rate at the detector is 
calculated by integrating the aperture over the neutrino flux and, if no
candidate events are found in a given time interval, a $90\%$ C.L.\ upper 
limit to the flux normalization is obtained matching the expected number of
detected events over the time interval to 2.39~\citep{Feldman:1997qc}. 
An assumption about the energy spectrum of the flux has to be made. It is
customary to again assume a canonical $E_{\nu}^{-2}$. In this 
case, $90\%$ of the detected events at the Observatory are between 100 PeV and
25 EeV~\citep{Aab:2015kma}. 
The most effective channel for neutrino detection is ES because the target 
mass provided by the Earth is large compared to the atmosphere and also 
because the threshold energy is low. For the ES channel algorithms 
have been found to identify events with just three triggered
stations. It has been shown that Earth-skimming events can only be observed
for zenith angles between $90^{\circ}$ and $95^{\circ}$, corresponding to
declination angles $-55^{\circ} < \delta < 60^{\circ}$. The exposure is
maximal for $\delta \simeq -53^{\circ}$ and $\delta \simeq 55^{\circ}$. The
apertures provided by the DGH and DGL are progressively smaller because, as
the zenith angle decreases, the atmosphere offers less target mass in which
the neutrinos can interact and be identified. 
As the search is effectively performed only for $\theta$ between
$60^\circ$ and $95^\circ$, the field of view of the Observatory for neutrinos
is limited to equatorial declinations between $\sim -85^\circ$ and $\sim
60^\circ$. 


When the pre-selected time interval is short, much less than a day, the 
aperture can be very different depending on the source position and the
observation time. The instantaneous aperture reaches
its maximum if the source position is just below the horizon with $\theta$
between $90^\circ$ and $95^\circ$. While the source is in the ES field of view,
the instantaneous aperture for neutrinos of energies above 100 PeV exceeds 
that of other neutrino telescopes. Also, depending on declination, the 
source can be inside the ES field of view for different lengths of time. 
For $\delta=0^\circ$ there are two transit periods per day into the ES field 
of view of about 25 minutes each. For the optimal declination positions 
($\delta \simeq -53^{\circ}$ and $\delta \simeq 55^{\circ}$), near the
extremes in declination of the ES band, the total transit time per day 
is more than four hours, as can be seen in Figs.~\ref{fig:fov_GW150914} 
(left) and \ref{fig:fov_GW170817}. 

The selection and identification procedures have been applied to the data
registered by the Observatory. As no neutrino candidates have been found,  
upper limits to the ultra-high energy (UHE) neutrino fluxes have been 
obtained. Upper limits to the diffuse flux with important implications 
for some models of UHECR production were first obtained in 
the ES channel~\citep{Abraham:2007rj,Abraham:2009pm} and subsequently  
extended to include the DGH~\citep{Abreu:2011zzd} and the
DGL~\citep{Aab:2015kma} channels. Upper limits to point source fluxes 
were obtained for the ES and DGH channels~\citep{Abreu:2012zz}. 
Upper limits on neutrino fluxes from point sources are shown in
Fig.\ref{fig:neutrino-search} (right) as a function of declination for the
search period up to 31 March 2017. Articles with further updates and details 
on both diffuse and point source searches are to appear soon. 

\subsection{Correlated searches of neutrinos}
\label{s:NuCorrelatedSearches}

\begin{figure}[t]
\centering
\includegraphics[width=.49\linewidth]{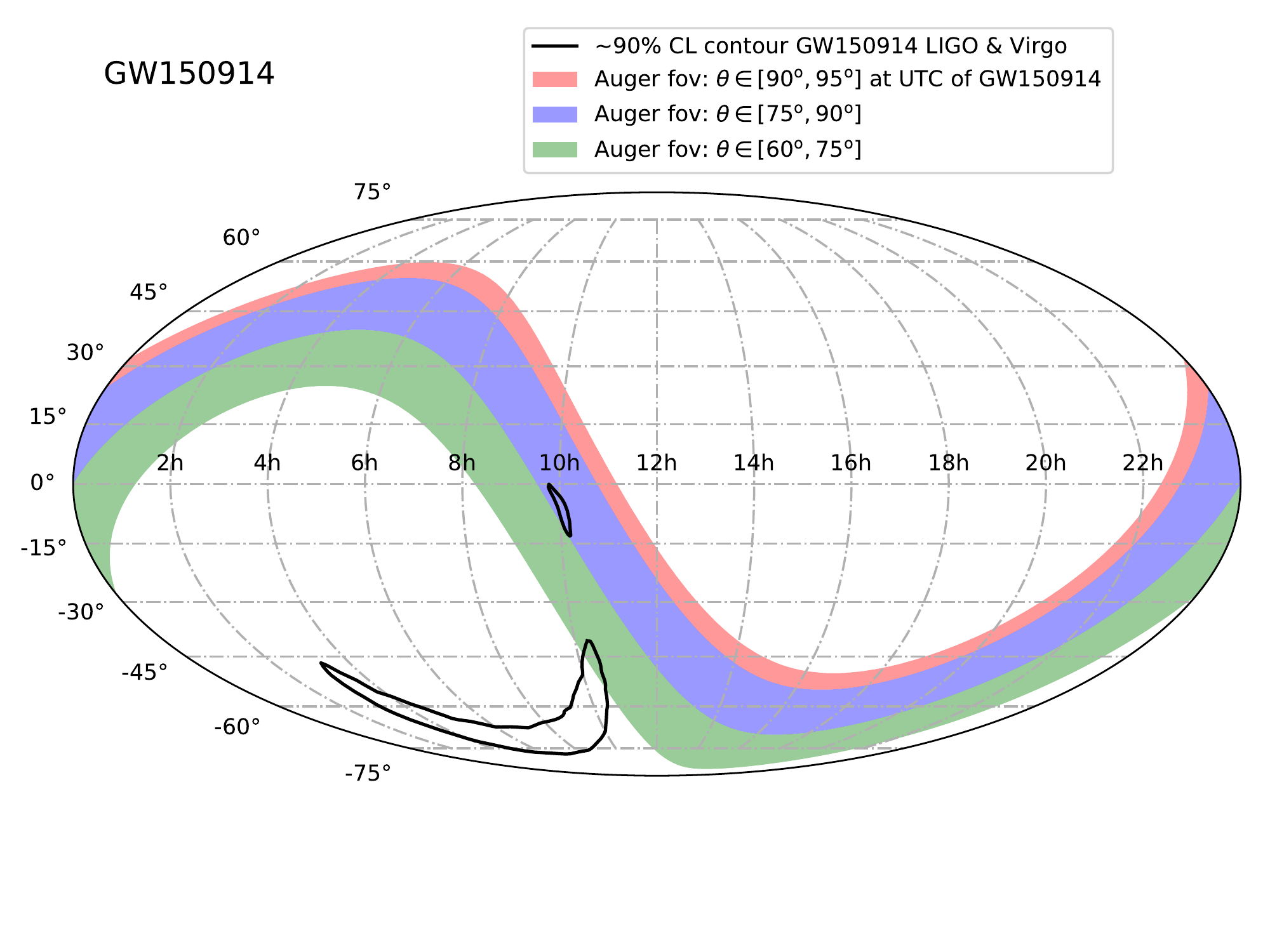}
\includegraphics[width=.49\linewidth]{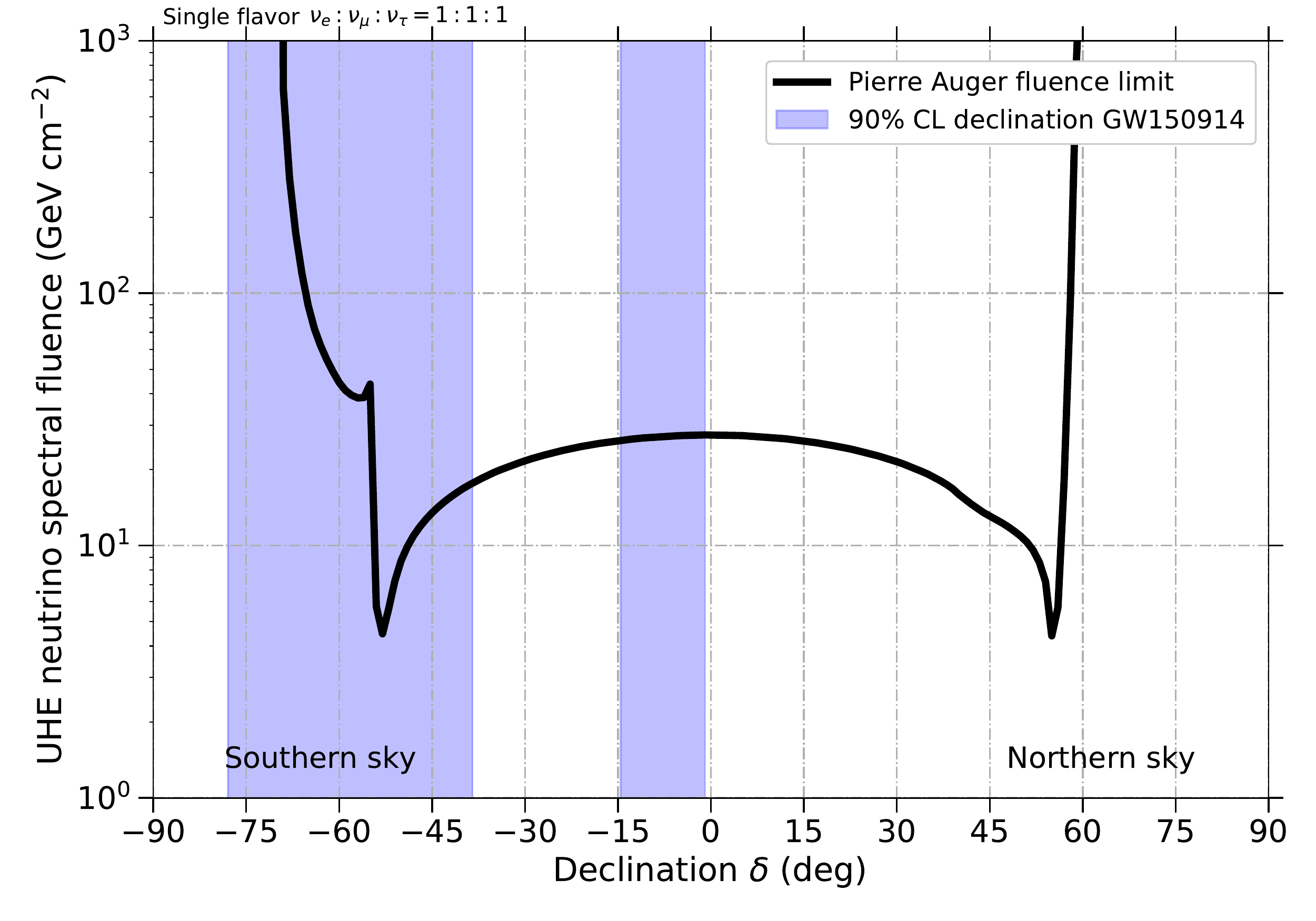}
\caption{Left: Field of view of the Observatory exemplified at the instant of
  detection of the black hole coalescence event GW150914 by Advanced LIGO
  \cite{LIGO-contours}. The band limits from top to bottom correspond to lines
  of zenith angles $95^\circ$, $90^\circ$, $75^\circ$ and $60^\circ$, used to
  separate the neutrino search into ES, DGH and DGL channels. The black
  contours give the $90\%$ C.L. region of the reconstructed position of the BH
  merger as obtained by LIGO observations. Right: Upper limit at $90\%$
  C.L. to the neutrino spectral fluence in the 100 PeV to 25 EeV range as a
  function of declination (see text), for the detection of black hole merger
  GW150914. The blue band is the $90\%$ C.L. of the reconstructed source
  declination, illustrating the limited precision level achieved using just
  detections at two LIGO sites.}

\label{fig:fov_GW150914}
\end{figure} 

The detection of gravitational waves from the merging of binary systems has
marked the birth of gravitational wave astronomy. The first signals from the
merging of two Black Hole (BH) systems GW150914~\citep{Abbott:2016blz} and 
GW151226~\citep{Abbott:2016nmj}
were reported in 2016. The luminosity distances of the two events were deduced
to be $410_{-180}^{+160}$ and $440_{-190}^{+180}$ Mpc and the radiated
energies $3.0_{-0.5}^{+0.5}$ and $1.0_{-0.2}^{+0.1}$ solar masses
respectively. They triggered the search for the emission of electromagnetic
radiation and neutrinos~\citep{Adrian-Martinez:2016xgn,Aab:2016ras}, even
though the black hole systems are not expected to emit any other type of
radiation unless matter debris and magnetic fields can be found in their
neighbourhood, possibly remaining from the BH
progenitors~\citep{Murase:2016etc,Kotera:2016dmp}.


Neutrinos were searched for 
with the Pierre Auger Observatory in two periods of time: $\pm 500$~s around
the UTC times at which the mergers were observed and one day following their 
occurrence~\citep{Aab:2016ras}. These intervals were motivated by the
association of the merging of compact objects to Gamma-Ray Bursts
(GRBs)~\citep{Moharana:2016xkz,Perna:2016jqh,Meszaros:2006rc}. Neutrinos have
been postulated to be produced by accelerated cosmic rays interacting with the
GRB photons in the prompt phase and with the lower energy photons of the
afterglow. The 1000~s time window is an upper bound of the duration of the
prompt GRB phase~\citep{ANTARES:2017iky,Kimura:2017kan} while the 1-day window
is a conservative bound of the duration of the
afterglow~\citep{Meszaros:2006rc}. 

\begin{figure}[tb]
\centering
\includegraphics[width=10cm]{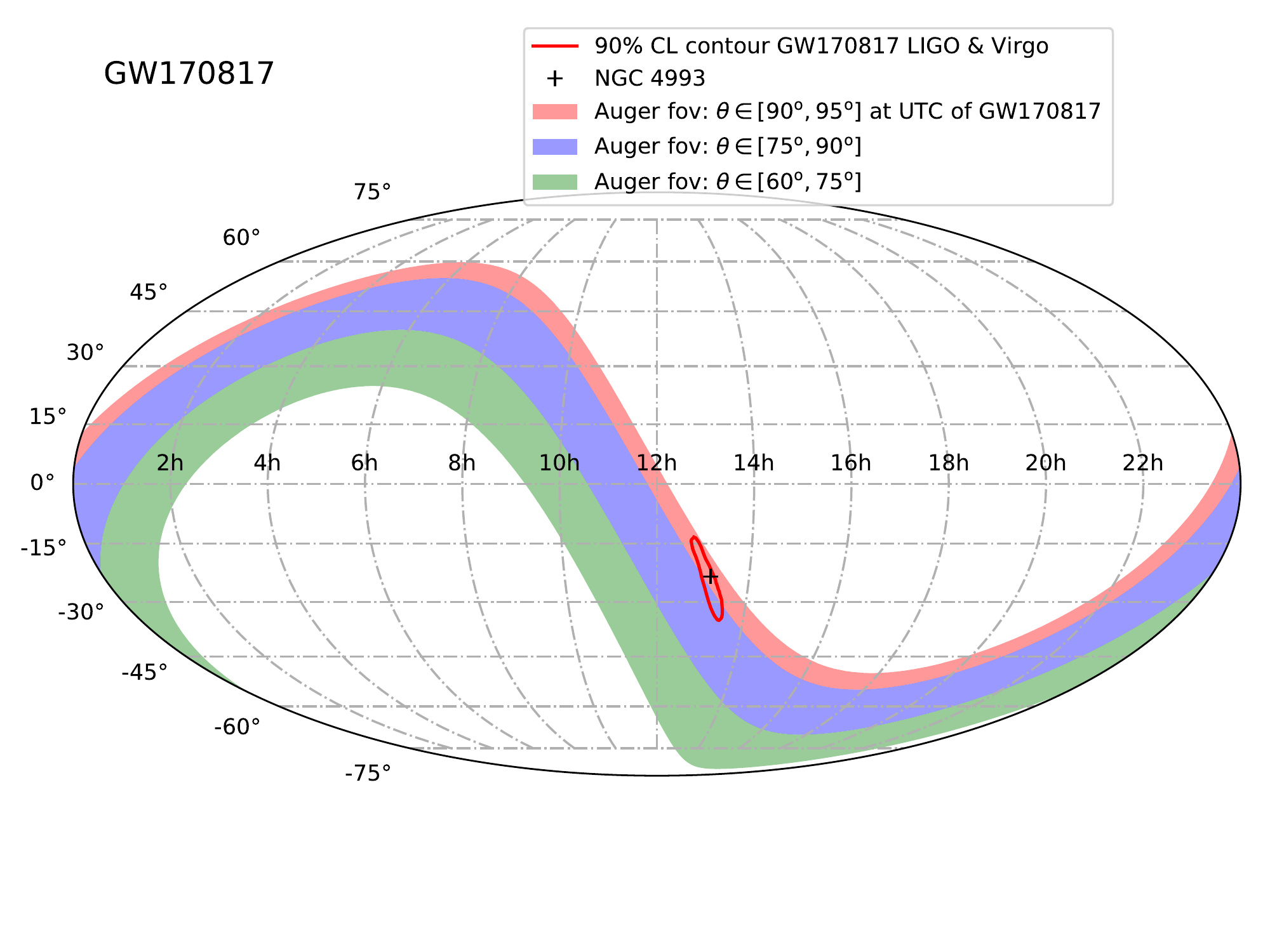}
\caption[]{Field of view of the Observatory in Earth-skimming and down-going
  channels at the instant of the detection of neutron star merger
  GW170817. The small red contour marks the event localization obtained by the
  Ligo-Virgo collaborations~\cite{LIGO-contours} and the black cross is the
  position of NGC\,4993, later correlated to the event by optical
  telescopes~\citep{Coulter:2017wya}.} 
\label{fig:fov_GW170817}
\end{figure} 
No neutrino candidates were observed in coincidence with these events in 
either of the time windows. 
Assuming a flux $k^{GW} E_\nu^{-2}$, where $k^{GW}$ is a
normalization constant, $90\%$ C.L.\ limits on the neutrino emission 
in the EeV from these events were reported as fluence limits. 
The spectral fluence is defined as $E_{\nu}^2 \phi(E_{\nu}) \Delta t$. Here $\phi$ is the spectral flux and $\Delta t$ is the chosen time interval 
over which the emission is assumed to take place and during which it is also 
assumed to be constant. In these conditions the spectral fluence is related 
to the total energy emitted in neutrinos. For the assumed spectrum it 
is trivial to convert the spectral fluence to an upper limit for the total 
energy emitted in neutrinos in a given energy band, which can be  
compared to the energy radiated in gravitational waves. The reported limits 
were obtained for a one-day period. Due to the poor localization of these
events the upper limits, shown in Fig.\ref{fig:fov_GW150914} (left) 
for GW150914, are reported as a function of declination. 
A similar limit was obtained for 
GW151226 since the distances to both mergers are quite similar. The most
restrictive limit assuming emission during a single day for GW150914 
(GW151226) is obtained for $\delta=-53^\circ$ ($\delta=55^\circ$) and would 
correspond to a total energy radiated in UHE neutrinos smaller than 
$7.7 \times 10^{53}$ ($9.7 \times 10^{53}$) erg when integrated from 100 PeV 
to 25 EeV. This corresponds to $14.3\%$ ($44.1\%$) of
the radiated energy in gravitational waves. 
The limits found by IceCube and ANTARES apply to lower neutrino energies and
give more restrictive limits on the total energy radiated in these
neutrinos~\citep{Adrian-Martinez:2016xgn}. 

Models have been proposed where 
black hole mergers accelerate UHECR~\citep{Kotera:2016dmp}, and the energy
budget of these collapsing events could account for all of the UHECR  with a
modest fraction of the gravitational wave energy (of order $3\%$) going to
UHECR acceleration. Assuming that a similar fraction goes into UHE neutrinos,
about 0.5 events could have been expected at the Observatory in coincidence
with GW150914 in the most optimistic scenario. The upper limits were
obtained averaging the instantaneous aperture over a day. If the emission 
time was shorter than one day, more stringent limits would be obtained provided the 
source was in the neutrino field of view during the neutrino emission, 
especially if it was in the ES field of view. Due to 
the poor localization of both events, it is not possible to know if the events 
were in the neutrino field of view at the detection time. Given the $90\%$ C.L. 
contour of GW151226 (GW150917) the overlap with the field of view of the neutrino 
search is $68.9\%$ ($13\%$) for a time window of 1000~s. The overlap over a whole 
sidereal day is $100\%$ in both cases. The detection of black hole mergers closer to 
us and with more directional precision could provide more stringent 
constraints for such models. 

\begin{figure}[tb]
\centering
\includegraphics[width=10cm]{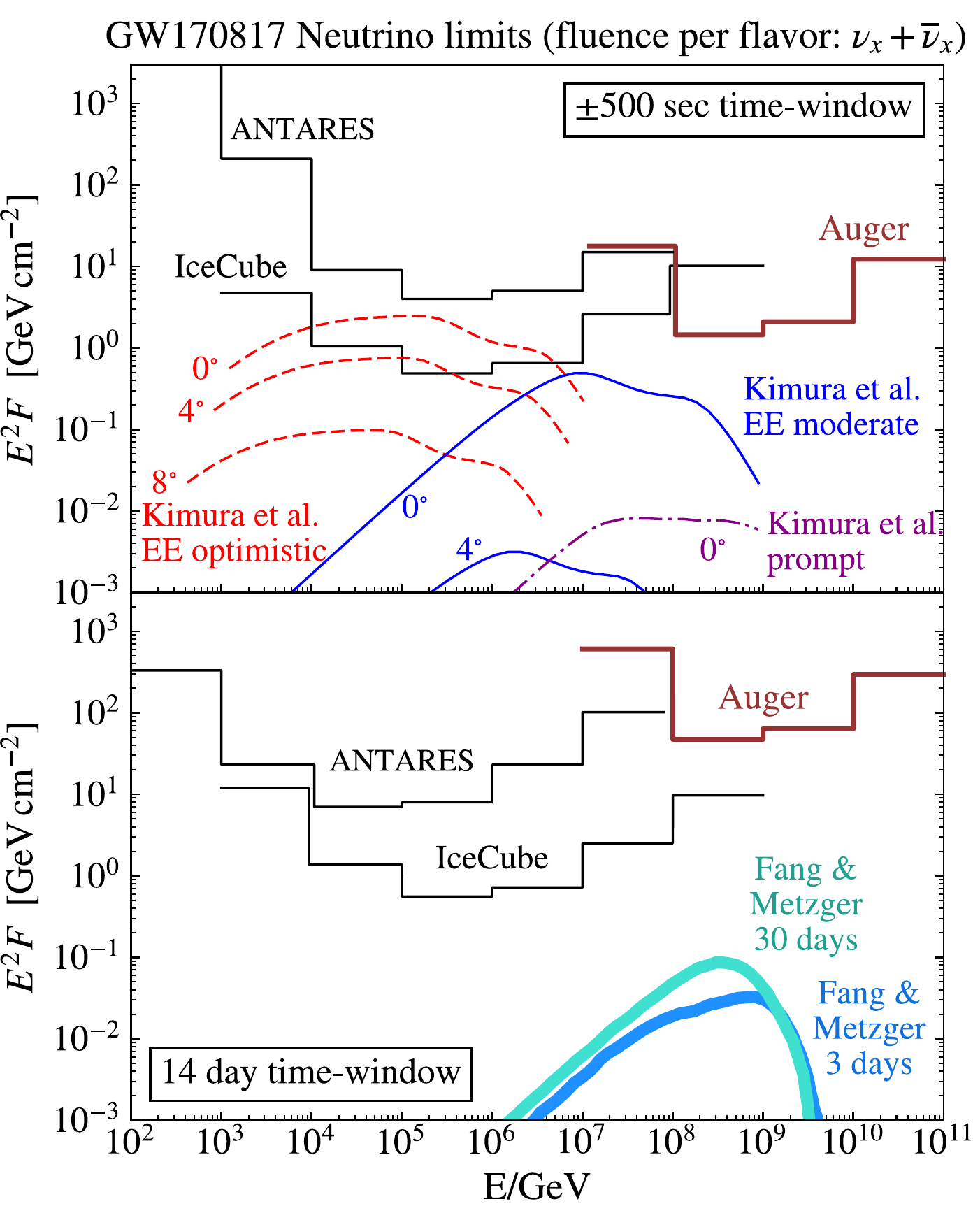}
\caption[]{Upper limits at $90\%$ C.L. on the neutrino spectral fluence from
  GW170817 for a  1000 s period centered at (14 day after) the time of the
  event in the top (bottom) panel. The bounds in the top panel are compared to
  predictions of models of prompt and extended emission
  (EE)~\citep{Kimura:2017kan} in the case of exact  alignment of the line of
  sight to the rotation axis and for selected off-axis viewing angles. The
  bounds obtained for a 14-day period on the bottom panel are compared to
  models of longer lived emission~\citep{Fang:2017tla}. All models have been
  scaled to 40 Mpc, the distance to the host galaxy NGC 4993. (Modified
  from~\citep{ANTARES:2017bia}).} 
\label{fig:GW170817FluenceBounds}
\end{figure} 
The observation of a neutron star merger GW170817 by the LIGO-Virgo
collaboration~\citep{Monitor:2017mdv} less
than two seconds before the short GRB 170817A detected by 
Fermi-LAT~\citep{Goldstein:2017mmi,Savchenko:2017ffs} 
from a coincident direction, has had outstanding implications. The succesful 
alert systems triggered observations and detection in practically all bands of 
the electromagnetic spectrum in an unprecedented way, marking the onset 
of a new era of Multimessenger Astronomy~\citep{GBM:2017lvd}. 
Early optical observations
allowed a precise localization slightly off-centered in the galaxy NGC 4993 at
equatorial coordinates $\alpha(J2000.0)=13^{\rm h}09^{\rm m}48^{\rm s}$,
$\delta(J2000.0)=-23^{\circ}22'53.''343$~\citep{Coulter:2017wya}. Searches for
neutrino events were also performed at neutrino telescopes and at the Pierre
Auger Observatory~\citep{ANTARES:2017bia}. At the time of the GW detection,
the source was located at a zenith angle of $91.9^\circ$ at the Observatory
site, just below the horizon and extremely close to the sweet-spot for
Earth-skimming neutrinos, see Fig.~\ref{fig:fov_GW170817}. When considered in
a time interval of $\pm 500$ s about the detection ($93.3^\circ < \theta <
90.4^\circ$),
the EeV exposure is larger than that of dedicated neutrino telescopes and
provides the most stringent upper limit to the neutrino fluence at 90\% C.L. in the 100 PeV to 25
EeV interval~\citep{ANTARES:2017bia}, complementary to IceCube and ANTARES, as
illustrated in   Fig.~\ref{fig:GW170817FluenceBounds}. The analysis was also
made for a longer time window of 14 days matching predictions from
longer-lived emission processes~\citep{Gao:2013mcx,Metzger:2016pju} and also
displayed in Fig.~\ref{fig:GW170817FluenceBounds}. 
Neutrino bounds obtained with the Pierre Auger Observatory and other neutrino
telescopes are within the range predicted in some models of GRB from neutron
star mergers. IceCube bounds allowed constraints to be placed on the
orientation angle relative to the jet axis (assumed to be coincident to the
rotation axis) for one of the most optimistic models.

Acceleration of hadrons in astrophysical objects inevitably leads to photon and neutrino production with similar energy fluences. On September 17th of 2017, a 290 TeV neutrino was detected at the IceCube telescope~\citep{Aartsen:2016nxy} in Antarctica and subsequent correlated gamma-ray observations pointed for the first time to a powerful blazar in a flaring state, TXS 0506+056, as a candidate source of the neutrino~\citep{IceCube:2018dnn}. This result was further supported by a neutrino burst from the same direction of the sky obtained in a time window of about 158 days when searching for time correlations from the same direction~\citep{IceCube:2018cha}. A search for possible coincident signals at the Auger Observatory was also performed when the results were made public. This source is $21\%$ of the sidereal day in the field of view of the neutrino search but it was not in the field of view at the exact time of the neutrino detection. No neutrinos were found~\citep{AlvarezMuniz:2018}. The details of this search and the bounds obtained are to be reported in a separate article. 

%% file: Photons.tex
\section{Photons}
\label{sec:photons}

\subsection{Photon identification}
\label{s:PhotonIdentification}

Ultra-high energy (UHE) photons are among the possible particles contributing to the flux of cosmic rays and may arise from a number of processes. Firstly, UHE photons are expected from the GZK-process just like UHE neutrinos and can be used to probe the parameter space of UHECR sources. Secondly, they may be produced by hadronic interactions of cosmic rays within the sources or within their local environment. In these cases, the photons are produced on average with around 10\,\% of the energy of the primary incident proton. Thirdly, a large flux of UHE photons has been predicted in top-down models with UHECR originating from the decay of supermassive particles such as topological defects or Super Heavy Dark Matter (SHDM) particles (see e.g.\ \citep{Bhattacharjee:1998qc,Aloisio:2015lva}). 

As opposed to neutrinos, photons undergo interactions with the extragalactic background light (EBL) inducing electromagnetic cascades. This process makes photons sensitive to the extragalactic environment (e.g.\ EBL and magnetic fields) but it also limits the volume from which EeV photons may be detected.  It is small compared to the GZK-sphere, but large enough to encompass the Milky Way, the Local Group of galaxies, and possibly Centaurus A, given an attenuation length of about 4.5 Mpc at EeV energies \citep{DeAngelis:2013jna}. The secondary electrons and positrons created by pair production in the photon fields can again interact with background photons via inverse-Compton scattering, resulting in an electromagnetic cascade that ends at GeV–TeV energies. Comparing these expected diffuse GeV-TeV fluxes measured with instruments sensitive in this energy range provides another important cross-link to $\gamma$-astronomy (see e.g.\ \citep{Batista:2016yrx}).

The search for UHE photon primaries is based on the different development and particle content of electromagnetic and hadronic air showers. The induced electromagnetic cascades develop slower than hadronic ones so that the shower maximum $X_{\mathrm{max}}$ is reached closer to the ground. Proton and photon simulated showers have average $X_{\mathrm{max}}$ values that differ by about 200 g\,cm$^{-2}$ in the EeV energy range. This difference is enhanced at energies above 10 EeV because of the Landau-Pomeranchuk-Migdal (LPM) effect \citep{Landau:1953um,Migdal:1956tc}. Above 50 EeV, photons also have a non-negligible probability to convert in the geomagnetic field \citep{Erber:1966vv,Mcbreen:1981yc,Homola:2006wf} producing a bunch of low-energy electromagnetic particles, called “pre-shower”, entering the atmosphere. The $X_{\mathrm{max}}$ of the pre-showered cascades is smaller than for non-converted ones and the separation between the average $X_{\mathrm{max}}$ for photons and proton correspondingly reduced. The fluorescence detector (FD) with its high resolution of about 15\,g\,cm$^{-2}$ and systematic uncertainties of less than about 8\,g\,cm$^{-2}$ for $X_{\mathrm{max}}$ \citep{Aab:2014kda,Bellido:2017cgf} is an ideal instrument to discriminate photon- from hadron-induced air showers with high sensitivity even in single events. Thus the FD is able to provide strong constraints on the photon fraction of UHECR.

\begin{figure}[t]
\centering
\includegraphics[width=.49\linewidth]{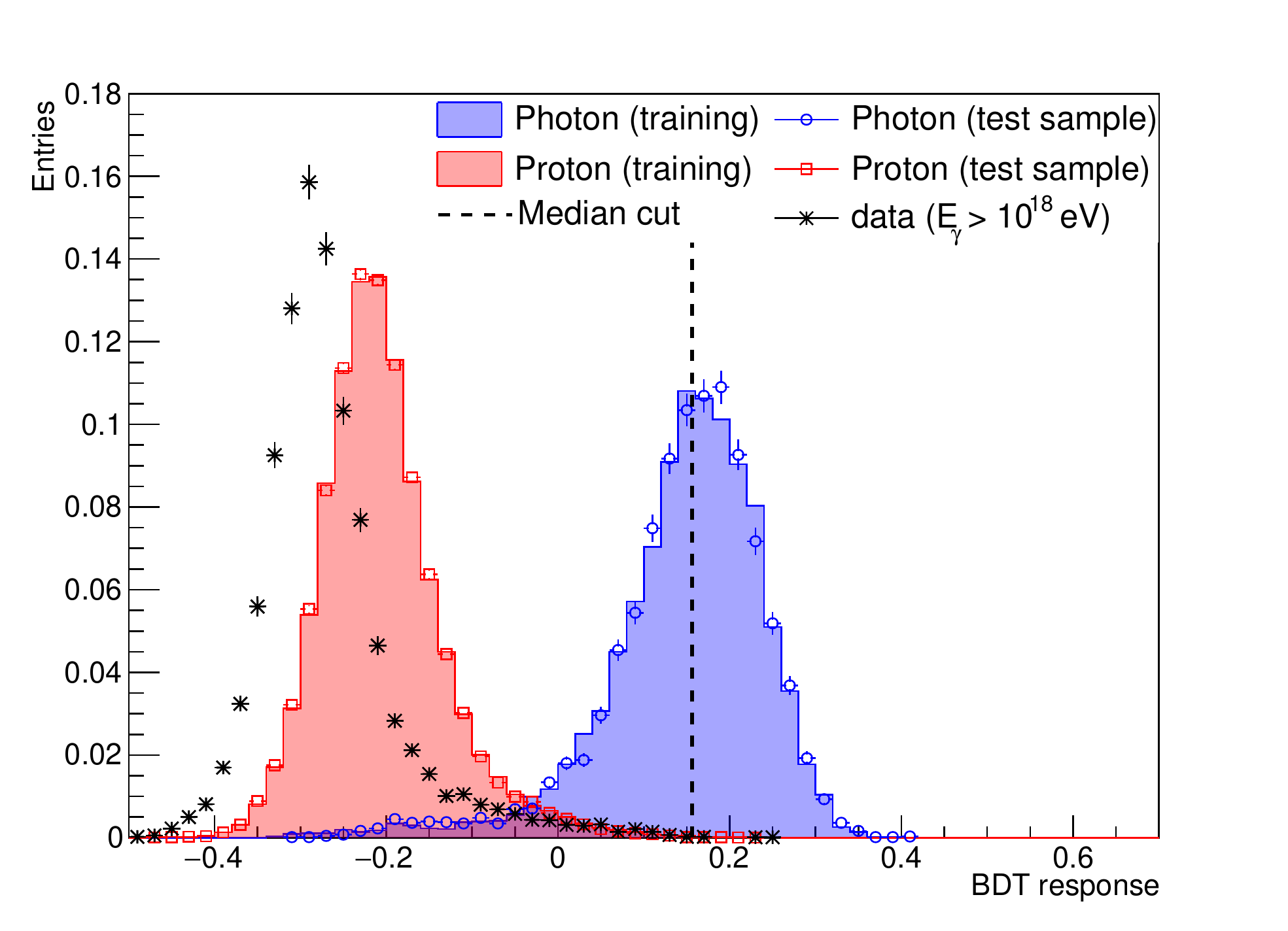}
\includegraphics[width=.49\linewidth]{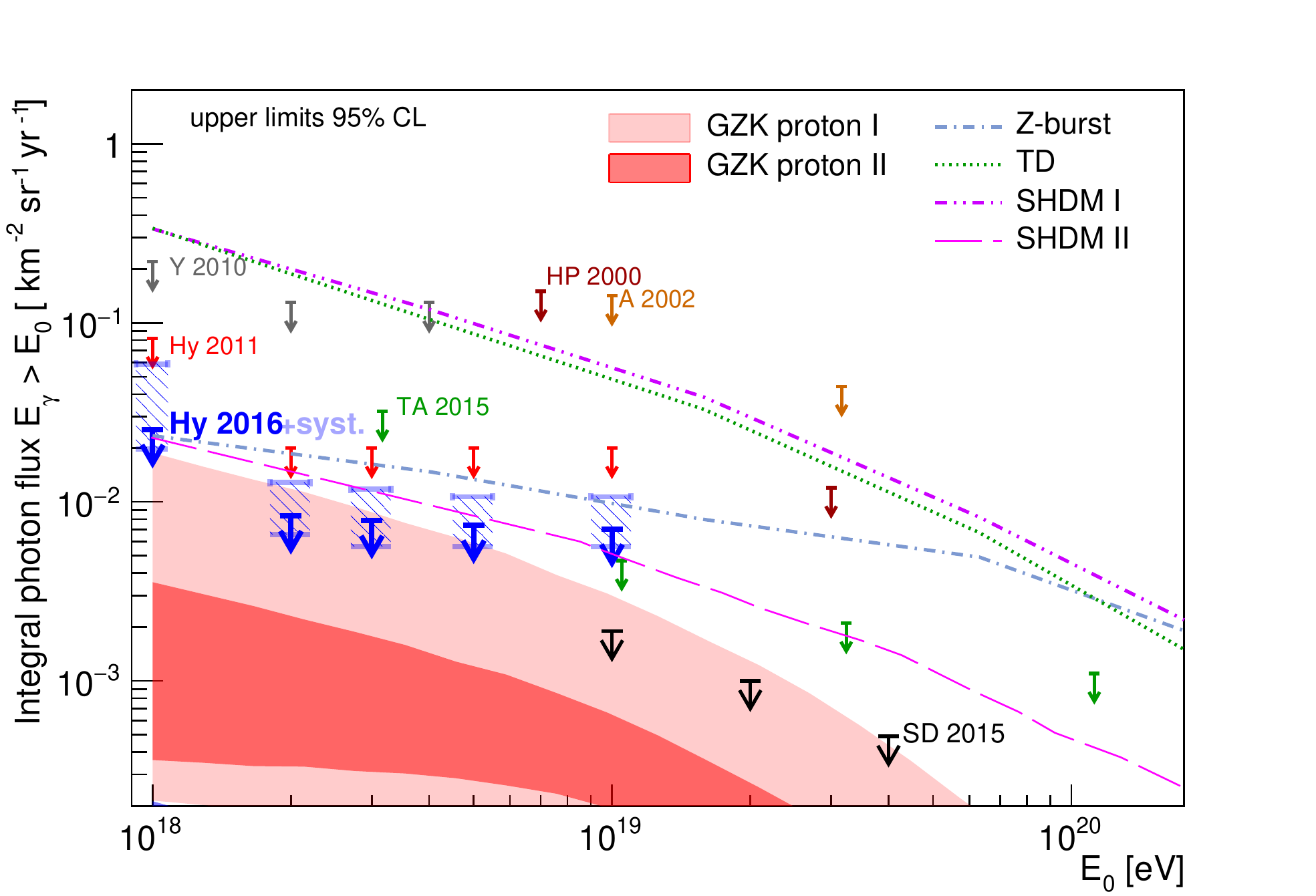}
\caption[]{Left: Photon identification with a Boosted Decision Tree for signal (photon, blue), background (proton, red) and data (black). For simulations, both the training and the test samples are shown. The cut at the median of the photon distribution is indicated by the dashed line. QGSJET-II-04 is used as high-energy hadronic interaction model.
Right: Compilation of upper limits on the integral photon fluxes from \citep{Aab:2016agp}. Blue arrows: Integral photon upper limits from the nine year hybrid data sample assuming a photon flux following $E^{-2}$ and with no background subtraction. The limits obtained when the detector systematic uncertainties are taken into account are shown as horizontal segments (light blue) delimiting a dashed-filled box at each energy threshold; Black arrows: Nine year SD data sample \citep{Bleve:2015nut}. Previous data from Auger as well as data from TA, AGASA, Yakutsk, and Haverah Park are included for comparison. The lines and shaded regions give the predictions for top-down models and GZK photon fluxes, respectively, assuming different parameters (references can be found in \citep{Aab:2016agp}).} 
\label{fig:photon-diffuse}
\end{figure} 

It is also possible to search for photon showers using the SD because the shower development and nature of the primary cosmic ray affect the content and the shape of the distribution of particles at the ground as a function of the distance from the shower axis (Lateral Distribution Function, LDF). Photon-induced showers generally are expected to have a steeper LDF compared to hadron primaries because of the sub-dominant role played by the flatter muonic component. The high-energy effects (LPM and pre-showering) affect the average muon content of air showers by typically less than 20-50\,\% dependent on zenith angle, which is small compared to the factor 5 to 8 difference in muons with regards to hadronic showers \citep{Risse:2007sd,Homola:2006wf}. However, the different stage of shower development at the ground (relative to  $X_{\mathrm{max}}$) leads to a modification of the observed LDF. Given the steeper LDF and the muon-driven SD triggers, the footprint at the ground, and consequently the number of triggered stations, $N_{\mathrm{stat}}$, is typically smaller for electromagnetic showers \citep{Abreu:2011zzd}. These features can be combined into a single observable $S_b$, that discriminates the photon- and hadron-induced air showers \citep{Aab:2016hkv}. Discrimination on a shower by shower basis is less efficient compared to the FD, but given the much larger statistics of the SD, strong constraints on the photon fluxes
can be provided. For hybrid observations, the discriminating observables $X_{\mathrm{max}}$,  $S_b$, and $N_{\mathrm{stat}}$ are injected into a Boosted Decision Tree (BDT) algorithm that takes into account also the energy and angular dependencies of the discriminating observables. To identify photons, a cut is defined at the median of the BDT response distribution for simulated photons, as depicted in Fig.\,\ref{fig:photon-diffuse} (left). We note that the discrepancy between the data and the proton simulations is in agreement with the current experimental indications of a composition varying from light to heavier in the EeV range \citep{Aab:2014aea} and with the muon deficit observed in simulations with respect to the Auger data \citep{Aab:2014pza,Aab:2016hkv}. Applying the cut in this conservative way, the signal efficiency remains constant independently of the composition and hadronic model assumptions. Events having a BDT response larger than the median cut are selected as “photon candidates”.  This yields a background contamination of $\sim 0.14$\,\% for proton showers using QGSJET-II-04. This background level overestimates the one expected in data because the composition is generally heavier than that of pure protons and because the interaction models underestimate the muon number in EAS, making the showers look more photon-like.  If the multivariate analysis is performed with a mixture of 50\,\% proton and 50\,\% iron as input to the training phase, the background contamination reduces to $\sim 0.04$\,\% with the main contribution coming from the smaller values of the simulated $X_{\mathrm{max}}$. At present, such probabilities are to be considered systematic uncertainty to the background of potential photon candidates, i.e.\ such background events would dilute the photon limits.

\subsection{Photon searches}
\label{s:PhotonSearches}

Figure \ref{fig:photon-diffuse} (right) shows, as a result, the upper limits on the integral photon flux derived from nine years (Jan.\ 2005 - Dec.\ 2013) of hybrid and SD data compared to results from other experiments \citep{Aab:2016agp}. Eight candidates were found in the first two energy intervals. These events are close to the applied photon cut and detailed simulations of hadronic showers will need to be performed to judge the probability that they are caused by UHECR rather than by photons. For now, they are conservatively considered background rather than signal so that the upper limits of the first two energy bins become less stringent. The differential flux limit between 10-30 EeV is found to be $6.80\cdot 10^{-11}$\,GeV\,cm$^{-2}$\,s$^{-1}$\,sr$^{-1}$ at 90\,\%\,C.L. Some top-down scenarios proposed to explain the origin of trans-GZK cosmic rays (dashed lines) are illustrated and rejected by previous bounds on the photon flux. A recent super-heavy dark matter proposal (SHDM II) developed in the context of an inflationary theory is shown as a long-dashed line \citep{Aloisio:2015lva}. Constraints on the lifetime-and-mass parameter space of SHDM particle will be imposed by current and future limits on the photon flux, as obtained for example in \citep{Kalashev:2016cre}. 


A prime interest in the search for photons is to identify the first UHE photon point sources or -- in case of non-observations -- to provide relevant upper limits thereby constraining the source characteristics. As already mentioned, the horizon is limited to only a few Mpc which reaches out to Cen\,A but not much farther \citep{Aab:2016bpi}.  If the energy spectra of TeV $\gamma$ sources measured by atmospheric Cherenkov telescopes \citep{Hinton:2009zz} extend to EeV energies, it is plausible that photons and neutrons from these sources are detectable also by the Auger Observatory. Sources that produce particle fluxes according to an $E^{-2}$ energy spectrum inject equal energy into each decade. Thus, a measured energy flux of 1\,eV\,cm$^{-2}$\,s$^{-1}$ in the TeV decade, as is found for a number of Galactic sources, would result in the same energy flux in the EeV decade if the spectrum continues to such high energies and energy losses during propagation are negligible. A source of specific interest is the galactic center for which the H.E.S.S.-collaboration measured a gamma-ray flux up to about 50 TeV without any observation of a cutoff or a spectral break, suggesting that the Galactic center hosts a peta-electron volt accelerator, called “PeVatron” \citep{Abramowski:2016mir}.

It is still debated whether these photons are produced in hadronic processes. An interesting test for this is provided by a search for neutrons from this direction, because neutrons would necessarily be produced in charge exchange interactions of protons at the source. This will be discussed in section \ref{sec:neutrons}. The ratio between photon and neutron emissivities from p–p collisions at the same primary proton energy depends primarily on the spectral index of the protons at the source. Assuming a continuation of the parent proton spectrum with a spectral index $\Gamma_p \lesssim 2.4$ well beyond the energies observed in the photon spectrum, the photon emissivities are expected to dominate over those of neutrons \citep{Crocker:2004bb}.

\begin{figure}[t]
\centering
\includegraphics[width=8cm]{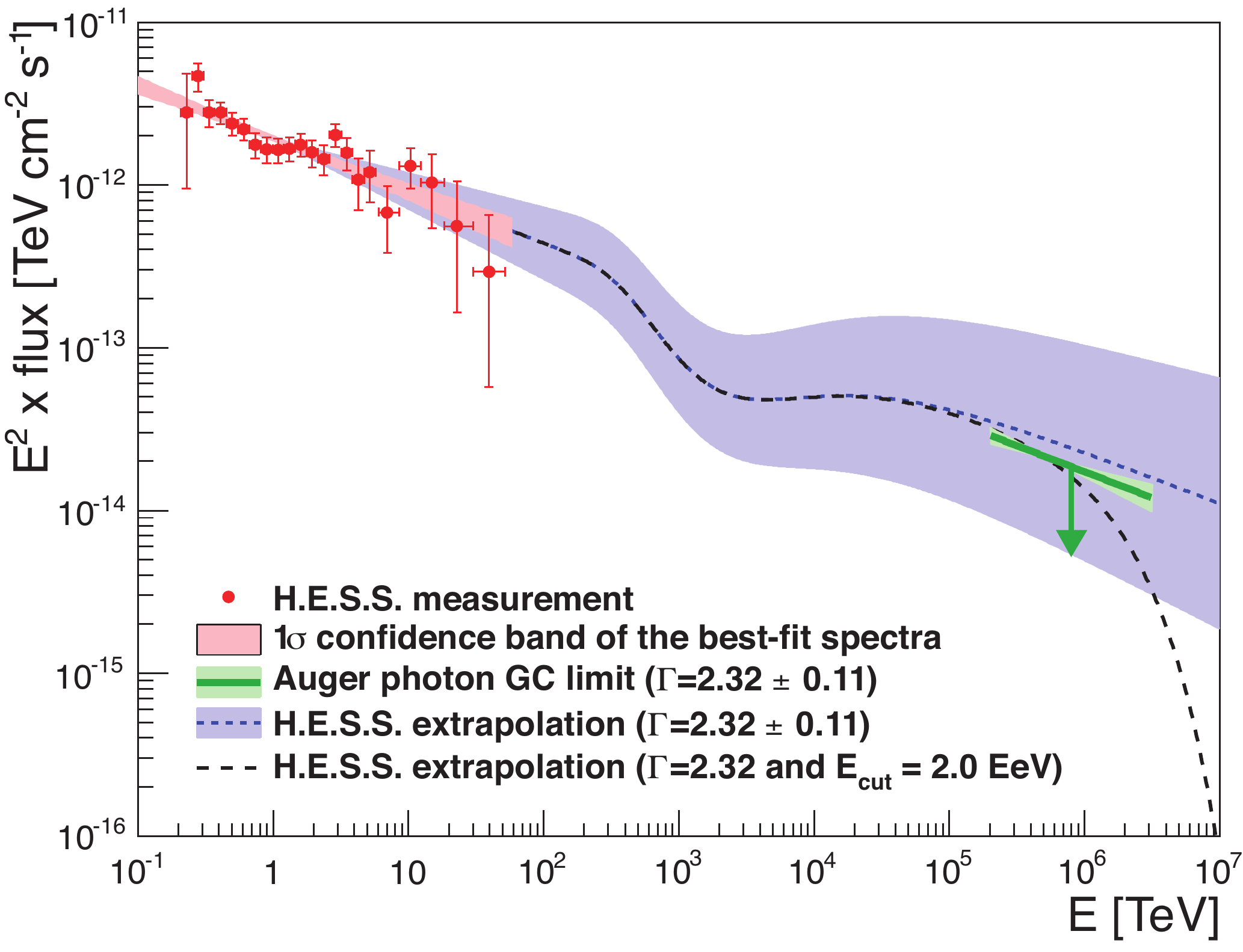}
\caption[]{Gamma-ray spectrum from the Galactic center region as measured by the H.E.S.S. collaboration (red points) \citep{Abramowski:2016mir}. The measured photon flux is extrapolated into the EeV range, given the quoted spectral index and its uncertainties (blue shaded region). The Auger limit \citep{Aab:2016bpi} is indicated by a green line (the green band reflects again the spectral uncertainties of the gamma-ray spectrum). A spectral index with cutoff energy $E_\mathrm{cut} = 2.0 \cdot 10^6$\,TeV is indicated by the dashed line.} 
\label{fig:photons-GC}
\end{figure} 

To search for photons from a list of predefined target sources \citep{Aab:2016bpi}, more relaxed cuts are applied to the observed events \citep{Aab:2014bha}.  For each candidate source direction an optimized cut in the multivariate output distribution is determined which depends on the expected number of isotropic background events in that direction. This number is calculated by applying a scrambling technique that takes into account detector efficiencies and aperture features. For each target direction, a top-hat counting region of $1^\circ$ is assumed \citep{Aab:2014bha}. Averaging over all considered target directions, the multivariate cut is expected to retain 81.4\,\% of photons while rejecting 95.2\,\% of background hadrons. In none of the sources and source classes could EeV photons be detected. As an example, the result for the Galactic center is illustrated in Fig.\,\ref{fig:photons-GC}. Conservatively, the extrapolation of the H.E.S.S. data to EeV energies does not take into account the increase of the p–p cross-section by about 40\,\% relative to TeV energies.  The current upper limit of 0.044 eV\,cm$^{-2}$\,s$^{-1}$ (95 \%\,C.L., $\Gamma=2$) at energies above 0.2\,EeV can already constrain the allowed parameter space for a flux continuation to EeV energies. 

The observation of photon fluxes from individual sources or from stacked sets of targets would have proved that EeV protons are being accelerated at those considered sources within the galaxy or its neighborhood. However, the null results leave open the possibility that protons observed at EeV energies are of extragalactic origin. This is in fact suggested by the large scale anisotropies observed with the Auger Observatory~\citep{Aab:2017tyv}. Moreover, the absence of detectable photon fluxes, as reported here, does not exclude the production of EeV protons within the Galaxy because the derived flux limits are time-averaged values. EeV photons might be produced in transient sources, such as gamma-ray bursts or supernovae, or be aligned in jets not pointing to us.  Extending the searches to bursting sources of EHE photons is a goal of ongoing multi-messenger analyses within Auger, briefly addressed in Section \ref{sec:Conclusions}.

%% file: Neutrons.tex
\section{Neutrons}
\label{sec:neutrons}


Like photons and neutrinos, neutrons travel in straight lines, undeflected by magnetic fields. They produce air showers that are indistinguishable from those produced by protons. Thus, a flux of neutrons from a discrete source would cause an excess of cosmic-ray events around the direction to the source, clustered within the angular resolution of the Observatory. Since free neutrons undergo $\beta$-decay with a mean lifetime of about 880\,s at rest, the mean travel distance for relativistic neutrons is 9.2 kpc$\times E_\mathrm{n}$/EeV. The distance from Earth to the Galactic center is about 8.3 kpc, and the radius of the Galaxy is approximately 15 kpc. Thus, sources in part of the Galactic disk, including the Galactic center, should be detectable via neutrons above 1 EeV. Above 2 EeV, the volume for detectable neutron emitters includes most of the Galaxy.

Neutron production is governed by charge exchange interactions of high energy cosmic-ray protons with ambient photons, protons, or nuclei leading to the creation of a $\pi^+$-meson. The $\pi^+$ takes the positive charge of the proton and a leading neutron emerges with most of the energy that the proton had.  The production of neutrons via creation of $\pi^+$-mesons is necessarily accompanied by photons originating from decay of similarly produced neutral pions. However, photons resulting from the decay of neutral pions acquire only a small fraction of the proton energy, so that the production of neutrons exceeds the hadronic production of photons of the same energy, provided the accelerated proton spectrum falls approximately like $1/E^2$ or more steeply with energy. This makes searches of high energy neutrons a highly relevant and sensitive probe for Galactic hadronic cosmic ray accelerators.

Similarly to the targeted search of EeV photon sources, discussed in Section \ref{sec:photons}, a flux of neutrons from astrophysical sources in the Galaxy can be detected in the Pierre Auger Observatory as an excess of cosmic-ray air showers arriving from the direction of the sources. To avoid the statistical penalty for making many trials, classes of objects were tested in combinations as “target sets”. Those target sets include msec pulsars (PSR), pulsar wind nebulae (PWN), microquasars, and magnetars (the full list of targets can be found in \citep{Aab:2014caa}).  In addition, a search for a neutron flux from the Galactic center and from the Galactic plane was performed. Within a target set, each candidate source is weighted in proportion to its electromagnetic flux, its exposure to the Auger Observatory, and its flux attenuation factor due to neutron decay. 

None of the searches provided evidence for a neutron flux from any class of candidate sources. Based on the first nine years of data (Jan.\ 1, 2004 to Oct.\ 31, 2013), the upper limits on the energy flux from these candidate sources, including the Galactic center, are mostly at a value of 0.10-0.15\,eV\,cm$^{-2}$\,s$^{-1}$ \citep{Aab:2014caa}, which is about a factor of 10 below the energy fluxes detected from TeV gamma-ray sources in the Galaxy \citep{Hinton:2009zz} and at about the level of sensitivity reached with the targeted photon searches, discussed above. If those sources were accelerating protons in the same environment to EeV energies with the $1/E^2$ dependence expected for Fermi acceleration, then the energy flux of neutrons in the EeV energy decade would even exceed the energy flux in TeV gamma rays.

Similarly, the energy flux of neutrons from the Galactic plane could be derived to less than 0.56\,eV\,cm$^{-2}$\,s$^{-1}$ (95\,\%\,C.L.). This provides a stringent constraint on models for continuous production of EeV protons in the Galaxy because the injection rate of protons into the Galactic disk must be sufficiently strong to balance their escape from the Galaxy. The concomitant neutron emission rate is model dependent. It could exceed the proton emission rate if protons are magnetically bound to the sources and only the produced neutrons escape, yielding EeV protons by their later decays. More likely, however, the neutron luminosity at any fixed energy is less than the proton luminosity. Based on an estimate of the proton emission rate, the limits on the neutron flux from the Galactic plane could be used to derive an upper limit on the ratio $\eta$ = neutron luminosity/proton luminosity. It was found to be $\eta_{\mathrm{UL}}\simeq 0.006$ \citep{Aab:2014caa}, which is a significant constraint on models for continuous production of EeV protons in the Galaxy.

%% file: Neutrino-UHECR-corr.tex
\section{Directional correlations of UHECRs with neutrinos and source candidates}


Even the most energetic cosmic rays can be subject to significant deflection and corresponding energy-dependent time delays. The strength and properties of the extragalactic magnetic field (EGMF) causing these effects remain largely unknown, and UHECR observed from bursting and continuous sources have already helped to provide constraints to the EGMF (see e.g.\ \citep{Lemoine:1997ei,Mollerach:2017idb,Bray:2018ipq}). In section \ref{s:NuCorrelatedSearches} we have discussed the time-correlated observation of neutrinos from merging binary systems. In fact, black hole mergers are also expected to produce UHECR and may provide sufficient luminosity to power the UHECR acceleration up to the highest energies \citep{Kotera:2016dmp}. On the other hand, 100 EeV protons from an extragalactic bursting source located at a distance of 30 Mpc are expected to arrive with time delays relative to photons and neutrinos by $\mathcal{O}(1-1000)$\,yrs \citep{Lemoine:1997ei} and may even reach out to $10^6$\,yrs in extreme cases, so that UHECR could not be included in those combined directional and temporal searches in a straightforward way. 

\subsection{Search for UHECR-Neutrino Correlations}
\label{s:uhecr-neutrino}

Purely directional correlations between high-energy neutrinos and UHECR still provide an interesting target of opportunity because UHECRs accelerated in astrophysical sources are naturally expected to produce high-energy photons and neutrinos in interactions with the ambient matter and radiation. 
If neutrinos result from the decays of pions produced in $p\gamma$ or $pp$ processes, they would carry about 3–5\,\% of the proton energy. Hence, neutrinos observed with energies of 30 TeV to 2 PeV, such as observed by IceCube, would have been produced by protons with energies in the 1 -- 100\,PeV range. 
These energies are much smaller than those registered by the Auger Observatory, and it is possible that the sources that produce the PeV neutrinos do not accelerate CRs up to ultra-high energies. In that case no correlations would be expected. On the other hand, the sources that produce the UHECR can also be expected to produce lower energy CRs. If the observed neutrinos come from the same sources as the UHECR, some degree of correlation in the arrival directions of the highest energy cosmic rays and the observed neutrinos could be expected depending on composition and on the strength of magnetic fields. 
It should also be noted that neutrinos can reach us from cosmological distances while the UHECR horizon is limited to only a few 100 Mpc. Thus, only a small fraction of the detected neutrinos could be expected to have originated from visible UHECR sources so that correlations, if they were found, cannot be very prominent. 

After some initial cross-correlation efforts by the ANTARES collaboration \citep{ANTARES:2012rcn,Palma:2017yvs} different analysis strategies have been devised in an intercollaborative effort of IceCube and Auger \cite{Aartsen:2015siv} that were more recently joined also by the ANTARES and Telescope Array (TA) collaborations \citep{Aartsen:2015dml,TheIceCube:2018gnp}. 

In a classical cross-correlation method, the number of observed UHECR-neutrino pairs is counted as a function of the maximal angular separation and then compared to the averaged background, which is calculated once for an isotropic neutrino flux and once for an isotropic UHECR flux. Applying this analysis to the latest combined Auger+TA UHECR and IceCube neutrino data sets, it was found that a maximum departure from the expectation for an isotropic CR flux, keeping the arrival directions of the neutrinos fixed, occurs at an angular distance of $1^\circ$ for tracks and $22^\circ$ for cascades, with post-trial p-values of $0.48$ and $5.4\cdot 10^{-3}$, respectively \citep{TheIceCube:2018gnp}.
In parallel, an unbinned likelihood method was used. It is based on a stacking point-source analysis applied to the UHECRs, where the neutrino arrival directions are smeared with a symmetric 2D-Gaussian and stacked as the signal template. The width of the Gaussian is the quadratic sum of the UHECR reconstruction uncertainty and the magnitude of the magnetic deflection, which is assumed to be described by $\sigma_{\mathrm{MD}} = 6^\circ/E_{\mathrm{CR}}$[100 EeV]. Again, the analysis is applied to track- and cascade-like high-energy neutrino candidates separately. Assuming an isotropic flux of neutrinos, the smallest post-trial p-values are found to be $2.2\cdot 10^{-2}$ and $1.7\cdot 10^{-2}$ for the track- and cascade-like events, respectively. 

These results are interesting, but the level of correlation is still too small to arrive at a conclusion. It will be interesting to see how the signal evolves as more data are accumulated by the  neutrino and UHECR observatories involved. In addition, new improved analysis strategies are being developed making, for example, better use of the measured UHECR composition and energy so that deflections in the EGMF and Galactic magnetic field can be more appropriately accounted for.

\subsection{Association of UHECR with Source Populations}
\label{s:uhecr-sources}

A related multi-messenger approach aimed at identifying the sources of UHECR from their directional information is to search for correlations with catalogue based astronomical objects. The basic concept is related to the studies presented in the previous section, only that the tracers are specific source populations rather than samples of high energy neutrinos. A set of source candidates that was studied in \citep{Aab:2018chp} includes nearby radio-loud AGNs and SBGs\footnote{We note that the potential of SGBs to accelerate particles to the highest energies has been questioned, recently \citep{Romero:2018mnb,Matthews:2018rpe}.}, selected as possible sources to accelerate CRs to the highest energies~\citep{Hillas:1985is,Dermer:2010iz}. 
Individual objects found within a given catalogue are expected to contribute differently to the total UHECR flux observed at Earth. Since the CR luminosity of individual sources is not known, we have chosen their observed electromagnetic emission as a proxy. In case of the 17 radio loud $\gamma$AGN found within 250 Mpc in the 2FHL catalogue from Fermi-LAT \citep{Ackermann:2015uya}, the measured integral $\gamma$-flux from 50\,GeV to 2\,TeV was used as a proxy for the UHECR flux. Similarly, 23 (out of a total of 63) SBGs within 250 Mpc and with a flux larger than 0.3 Jy were taken from \citep{Ackermann:2012vca} and weighted with their continuum radio emission at 1.4\,GHz. In both cases, attenuation of UHECRs from distant objects in the photon background fields was accounted for using a data-driven scenario that reproduces the composition and spectral constraints obtained by the Observatory \citep{Aab:2016zth}. Within this Ansatz, the relative contributions of NGC\,4945, NGC\,253, and M83 are expected to be about 38, 33, 13\,\% of the total UHECR flux from SBGs observed at the Auger Observatory. The  relative contributions seen at the Observatory from the $\gamma$AGN would be dominated by the Cen A Core (75\,\%), followed by M87 (15\,\%) and Mkn\,421  (8\,\%).

\begin{figure}[tb]
\centering
\includegraphics[width=.45\linewidth]{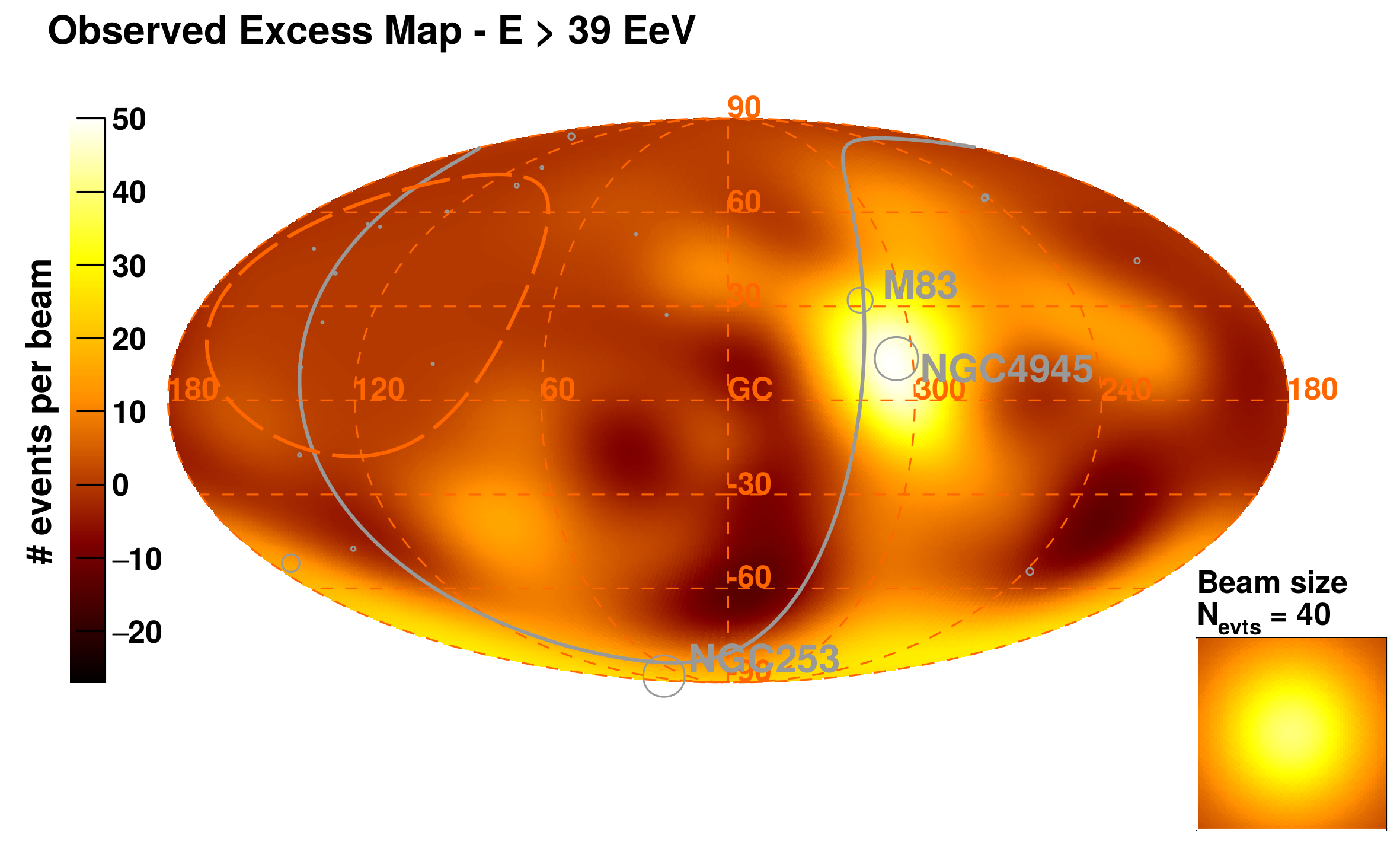}
\includegraphics[width=.45\linewidth]{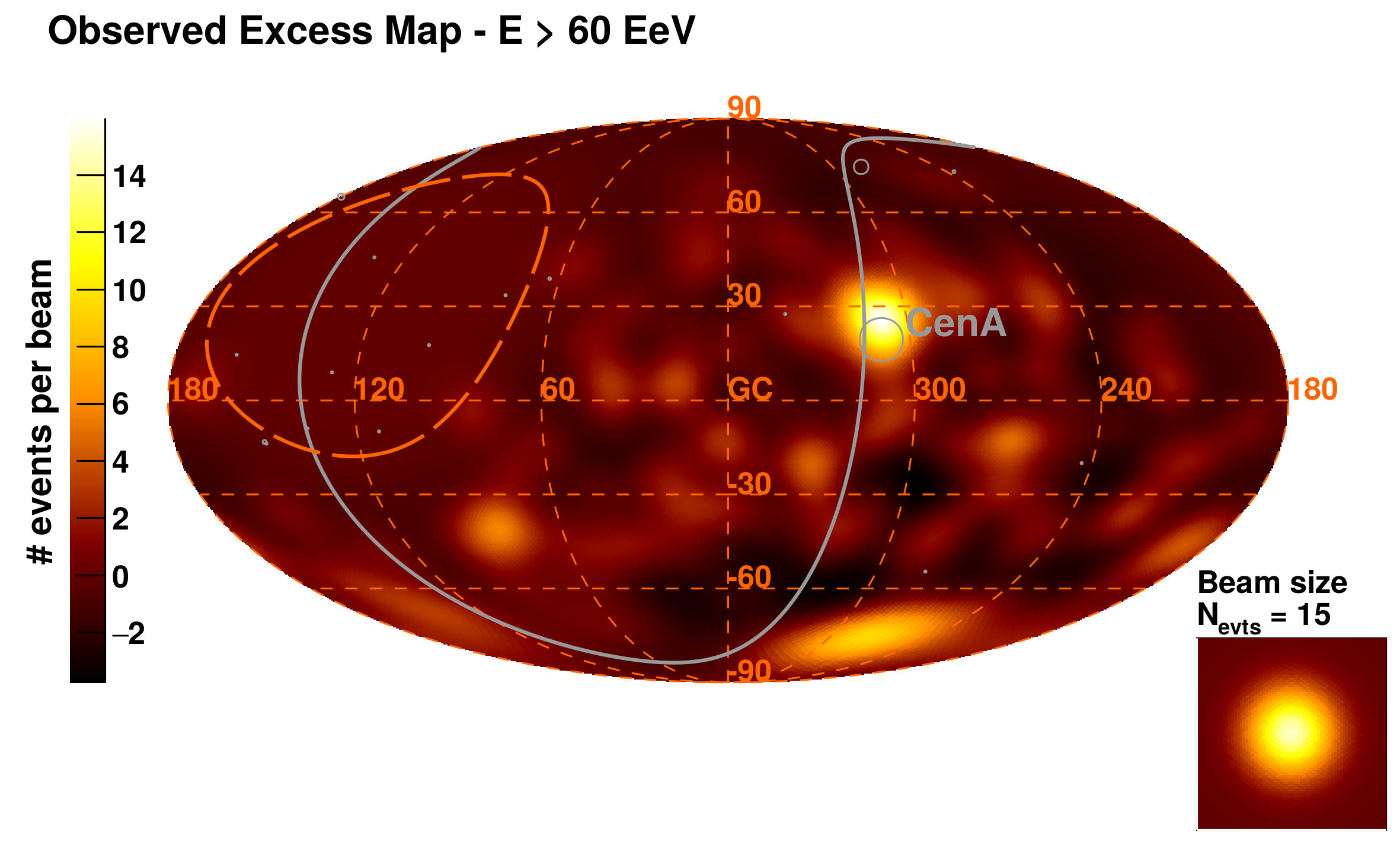}
\centering
\includegraphics[width=.45\linewidth]{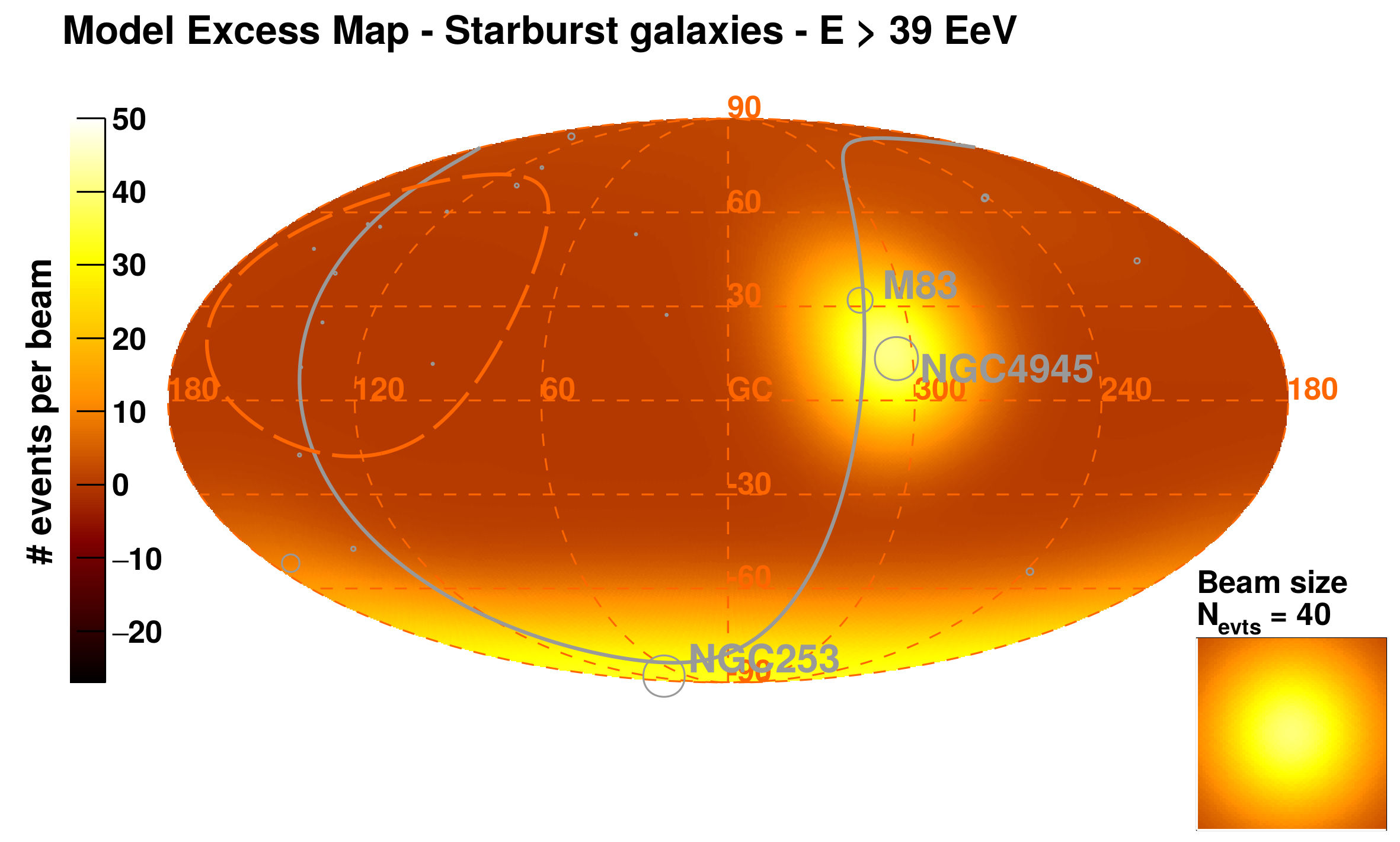}
\includegraphics[width=.45\linewidth]{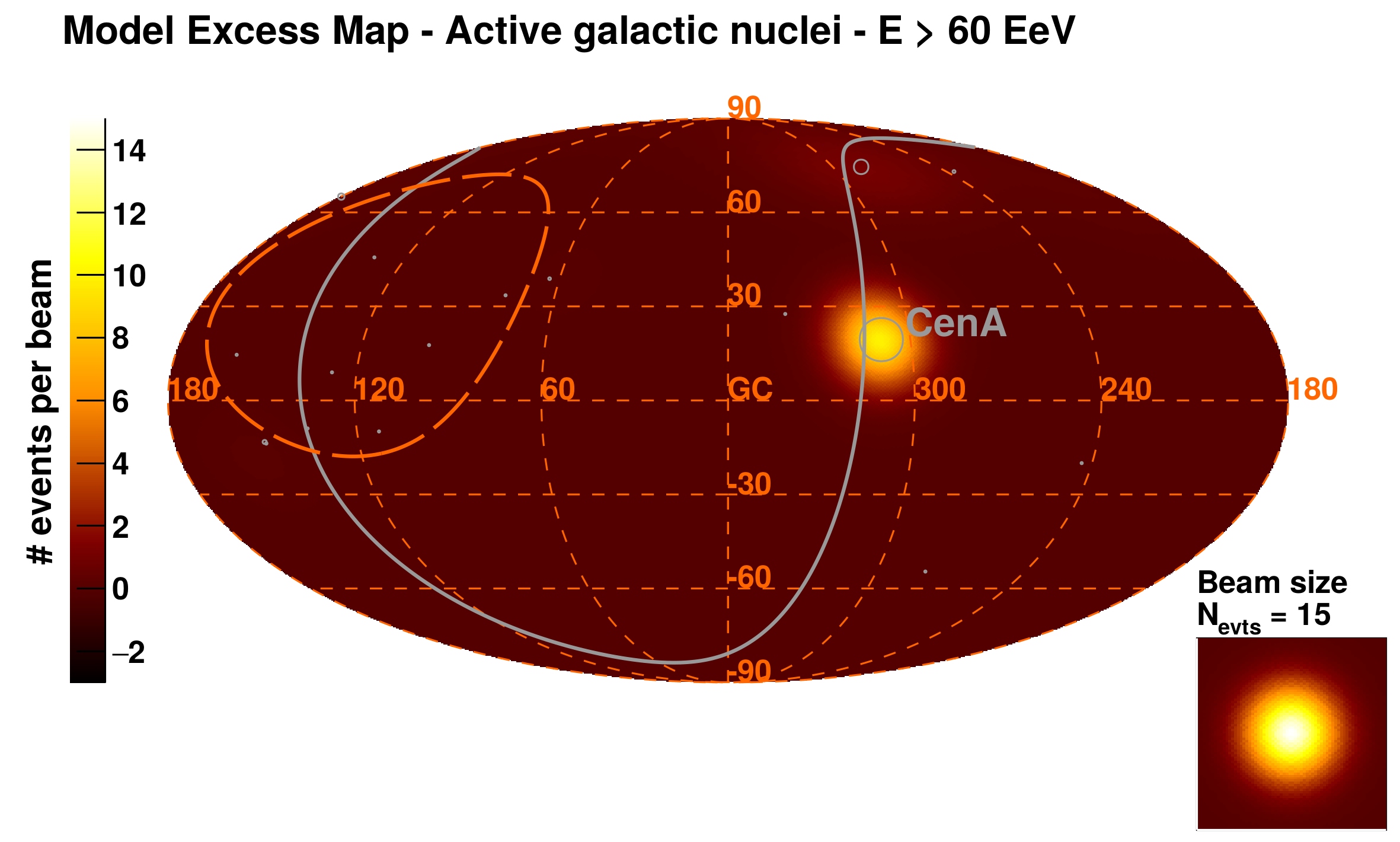}
\caption{Top row: observed excess maps for $E_{\mathrm{CR}}>39$\,EeV
and $E_{\mathrm{CR}}>60$\,EeV.
Bottom row: model excess maps for the SGB (left) and $\gamma$AGN model (right) at the same energy thresholds as above. The color scales indicate the number of events per smearing beam (see inset). The model flux map corresponds to a uniform full-sky exposure. The supergalactic plane is shown as a solid gray line. An orange dashed line delimits the field of view of the array. (From \citep{Aab:2018chp}).} 
\label{fig:sbg-agn-maps}
\end{figure} 

To test these models against the observed UHECR sky maps, a maximum-likelihood analysis was performed with two free parameters aimed at maximizing the degree of correlations of the model maps with UHECR events: a) the fraction of an isotropic component contributing to the total UHECR flux in addition to the source population being tested, and b) the width of a 2D Gaussian smearing around the position of the source candidates. The first free parameter can be interpreted as the contribution from more distant or less luminous sources, and the second parameter effectively accounts for the random scattering of UHECR in the EGMF. This analysis is repeated for a sequence of energy thresholds applied to the UHECR events and the test statistic (TS) as a function of the two parameters is analyzed. The results are shown in Fig.\,\ref{fig:sbg-agn-maps}. It was found that the SBG-Model rejects the hypothesis of an isotropic sky best at $E_{\mathrm{CR}}>39$\,EeV and yields a post-trial significance of $4.0\sigma$. Here, the SBGs contribute about 10\,\% to the total flux, and the smearing angle is $12.9^\circ$. In the case of the $\gamma$AGN model, the best rejection of isotropy is reached at $E_{\mathrm{CR}}>60$\,EeV at a level of $2.7\sigma$ with a $\gamma$AGN fraction to the total flux of about 7\,\% and a smearing angle $6.9^\circ$. Interestingly, the smearing angle decreases with increasing energy as expected from the larger rigidities of the UHECR in the $\gamma$AGN model. 

The results are very promising and demonstrate departures from a pure isotropic sky with more and more structure becoming visible, starting at above 8 EeV with the observation of a large-scale dipole \citep{Aab:2017tyv}, indications of higher order moments above 16 EeV, \citep{Aab:2018mmi}, blurry spots above 40 EeV, and hints of sharper spots above 60 EeV.
There are indications that both SGB and AGN, as well as a mixture of both or similar populations could  play a significant role in the production of UHECR. However, it is still too early to conclude about this long standing problem.
Future analyses, possibly profiting from full-sky observations with TA, may involve a better modelling of the Galactic magnetic field, account for the change of the composition as a function of energy, and possibly introduce better proxies for the unknown UHECR luminosities, so that $5\sigma$ detections are well within reach.

%% file: Conclusions.tex
\section{Conclusions and Prospects}
\label{sec:Conclusions}

The Pierre Auger Observatory was designed as a multi-purpose observatory for the high-energy Universe with multi-messenger observations being foreseen from the beginning. In this review, we have summarized a few prominent examples that demonstrate the unprecedented sensitivities to UHE photons, neutrinos, and neutrons. Observations of photons in the so far dark Universe at $E_\gamma \gtrsim 10^{17}$\,eV or of a neutrino at these energies would be a major breakthrough by itself. However, the non-observation of point sources and diffuse fluxes of photons and neutrinos allowed the derivation of upper bounds that constrain models very effectively. The bounds to diffuse photon and neutrino fluxes have ruled out the top-down models of UHECRs origin motivated by particle physics and also started to constrain the GZK-effect as a dominant process for explaining the observed flux suppression of the most energetic UHECR~\cite{Zas:2017xdj}.  Many TeV $\gamma$-sources are observed at energy fluxes of the order of 1\,eV\,cm$^{-2}$\,s$^{-1}$. Such sources would be visible to the Auger Observatory as strong photon and Galactic neutron sources if their energy spectrum would continue with a Fermi-like energy distribution up to about $10^{17}$\,eV. Again, their absence suggests that their maximum source energy does not reach out to the threshold energy of the Observatory and/or that their spectrum is significantly softer than $dJ/dE \propto E^{-2}$.

Point source searches now routinely include also mergers of compact binaries alerted by gravitational wave interferometers. The most spectacular event so far was the neutron star merger GW170817 at a distance of about 40 Mpc. Within the predefined $\pm 500$\,s search window, the Auger Observatory reached a neutrino flux sensitivity above 100 PeV that was over an order of magnitude higher than of any other neutrino observatory presently operated. Again, the absence of neutrinos at Auger, IceCube and ANTARES allowed constraining the jet properties of the neutron star merger. The third observation run O3 is about to start with many more events being expected in the near future. To accommodate for this, mechanisms are set up to automatically react to GW or other astrophysical alerts and search for neutrinos and photons.

While receiving alerts from a worldwide network of observatories, it is also possible to send alerts from the Observatory. The Auger Observatory is both a triggering and a follow-up partner in the Astrophysical Multi-messenger Observatory Network (AMON) \citep{Smith:2012eu}, which establishes and distributes alerts for immediate follow-up by subscribed observatories with private or Gamma-Ray Coordinates Network\footnote{\url{https://gcn.gsfc.nasa.gov}} notices. AMON registers all vertical ($\theta \leq 60^{\circ}$), high-quality, Auger SD events, with energy above 3 EeV with a cadence of at most a few minutes. AMON establishes clusters of two or more Auger events received correlated in time and arrival direction as alerts. These are excesses of events expected from neutral  (photon or neutron) UHECRs as discussed in Section \ref{sec:photons} and Section \ref{sec:neutrons}. Unlike the aforementioned neutron and photon searches, this alert channel is sensitive to ultra-high energy neutral transient emission which is less well constrained in the time-integrated searches. For each Auger alert, AMON also receives and transmits the background rate, in other words, the significance of each of the alerts, which depends on the energy and declination of each event. Thus high-significance alerts can be preferentially followed-up. To improve the overall significance of alerts and to enhance future photon and neutrino searches in general,  additional improved hardware triggers have been implemented to the existing surface detector array of the Observatory. Besides enhancing the sensitivity to photons and neutrinos, they allow reducing their detection thresholds. Thus, photon alerts for individual ultra-high energy photon candidates from a selection of events which utilizes observables based on the new triggers will be transmitted in near-real time and available for immediate follow-up using the AMON infrastructure. Finally, the infilled array, covering an area of 23.5\,km$^2$ with a 750\,m detector spacing, will provide acceptance for neutrinos and photons below the present detection threshold, though with reduced total exposure. It may also be used to generate specific alerts and there are plans to extend the neutrino search to the FD~\cite{Aramo:2004pr}. 

Presently, the Pierre Auger Observatory is being upgraded to AugerPrime primarily to improve the mass composition measurements and particle physics capabilities with the surface detector array ~\citep{Aab:2016vlz,Engel:2015ibd,Martello:2017bqq}. For this purpose, 3.8\,m$^2$ plastic scintillation detectors will be installed on each of the existing surface detector stations and the current readout electronics will be replaced by a faster and more powerful one. The new electronics will facilitate three times faster sampling of the signals (120\,MHz instead of 40\,MHz), will have more monitoring features implemented, and will have more sensitive triggers installed for low-energy showers and those initiated by photons and neutrinos. Moreover, the infilled array will be furnished with large area underground muon detectors~\citep{PierreAuger:2016fvp,Aab:2017hhe} and each surface detector station will be amended by a radio antenna to provide improved shower information for inclined air showers~\citep{Horandel:2018}. AugerPrime will be operated into 2025 and will improve the statistics for composition information and composition enhanced astronomy by about a factor of 10. Clearly, all of these features will open up many new possibilities for improved searches of photons and neutrinos. This suit of enhancements will further strengthen the prominent role of the Pierre Auger Observatory as a multi-messenger observatory for the next decade.

%% file: acknowledgments.tex

\subsection*{Acknowledgments}

\begin{sloppypar}
The successful installation, commissioning, and operation of the Pierre
Auger Observatory would not have been possible without the strong
commitment and effort from the technical and administrative staff in
Malarg\"ue. We are very grateful to the following agencies and
organizations for financial support:
\end{sloppypar}

\begin{sloppypar}
Argentina -- Comisi\'on Nacional de Energ\'\i{}a At\'omica; Agencia Nacional de
Promoci\'on Cient\'\i{}fica y Tecnol\'ogica (ANPCyT); Consejo Nacional de
Investigaciones Cient\'\i{}ficas y T\'ecnicas (CONICET); Gobierno de la
Provincia de Mendoza; Municipalidad de Malarg\"ue; NDM Holdings and Valle
Las Le\~nas; in gratitude for their continuing cooperation over land
access; Australia -- the Australian Research Council; Brazil -- Conselho
Nacional de Desenvolvimento Cient\'\i{}fico e Tecnol\'ogico (CNPq);
Financiadora de Estudos e Projetos (FINEP); Funda\c{c}\~ao de Amparo \`a
Pesquisa do Estado de Rio de Janeiro (FAPERJ); S\~ao Paulo Research
Foundation (FAPESP) Grants No.~2010/07359-6 and No.~1999/05404-3;
Minist\'erio da Ci\^encia, Tecnologia, Inova\c{c}\~oes e Comunica\c{c}\~oes (MCTIC);
Czech Republic -- Grant No.~MSMT CR LTT18004, LO1305, LM2015038 and
CZ.02.1.01/0.0/0.0/16\_013/0001402; France -- Centre de Calcul
IN2P3/CNRS; Centre National de la Recherche Scientifique (CNRS); Conseil
R\'egional Ile-de-France; D\'epartement Physique Nucl\'eaire et Corpusculaire
(PNC-IN2P3/CNRS); D\'epartement Sciences de l'Univers (SDU-INSU/CNRS);
Institut Lagrange de Paris (ILP) Grant No.~LABEX ANR-10-LABX-63 within
the Investissements d'Avenir Programme Grant No.~ANR-11-IDEX-0004-02;
Germany -- Bundesministerium f\"ur Bildung und Forschung (BMBF); Deutsche
Forschungsgemeinschaft (DFG); Finanzministerium Baden-W\"urttemberg;
Helmholtz Alliance for Astroparticle Physics (HAP);
Helmholtz-Gemeinschaft Deutscher Forschungszentren (HGF); Ministerium
f\"ur Innovation, Wissenschaft und Forschung des Landes
Nordrhein-Westfalen; Ministerium f\"ur Wissenschaft, Forschung und Kunst
des Landes Baden-W\"urttemberg; Italy -- Istituto Nazionale di Fisica
Nucleare (INFN); Istituto Nazionale di Astrofisica (INAF); Ministero
dell'Istruzione, dell'Universit\'a e della Ricerca (MIUR); CETEMPS Center
of Excellence; Ministero degli Affari Esteri (MAE); M\'exico -- Consejo
Nacional de Ciencia y Tecnolog\'\i{}a (CONACYT) No.~167733; Universidad
Nacional Aut\'onoma de M\'exico (UNAM); PAPIIT DGAPA-UNAM; The Netherlands
-- Ministry of Education, Culture and Science; Netherlands Organisation
for Scientific Research (NWO); Dutch national e-infrastructure with the
support of SURF Cooperative; Poland -- National Centre for Research and
Development, Grant No.~ERA-NET-ASPERA/02/11; National Science Centre,
Grants No.~2013/08/M/ST9/00322, No.~2016/23/B/ST9/01635 and No.~HARMONIA
5--2013/10/M/ST9/00062, UMO-2016/22/M/ST9/00198; Portugal -- Portuguese
national funds and FEDER funds within Programa Operacional Factores de
Competitividade through Funda\c{c}\~ao para a Ci\^encia e a Tecnologia
(COMPETE); Romania -- Romanian Ministry of Research and Innovation
CNCS/CCCDI-UESFISCDI, projects
PN-III-P1-1.2-PCCDI-2017-0839/19PCCDI/2018, PN-III-P2-2.1-PED-2016-1922,
PN-III-P2-2.1-PED-2016-1659 and PN18090102 within PNCDI III; Slovenia --
Slovenian Research Agency; 
Spain -- 
Ministerio de Econom\'\i a, Industria y Competitividad (FPA2017-85114-P and FPA2017-85197-P), Xunta de Galicia (ED431C 2017/07),
> Junta de Andaluc\'{\i}a (SOMM17/6104/UGR), Feder Funds, RENATA Red Nacional Tem\'atica de Astropart\'\i culas (FPA2015-68783-REDT) and
> Mar\'\i a de Maeztu Unit of Excellence (MDM-2016-0692); 
USA -- Department of
Energy, Contracts No.~DE-AC02-07CH11359, No.~DE-FR02-04ER41300,
No.~DE-FG02-99ER41107 and No.~DE-SC0011689; National Science Foundation,
Grant No.~0450696; The Grainger Foundation; Marie Curie-IRSES/EPLANET;
European Particle Physics Latin American Network; European Union 7th
Framework Program, Grant No.~PIRSES-2009-GA-246806; and UNESCO.
\end{sloppypar}

%% file: latex_authorlist_authors.tex
A.~Aab$^{75}$,
P.~Abreu$^{67}$,
M.~Aglietta$^{50,49}$,
I.F.M.~Albuquerque$^{19}$,
J.M.~Albury$^{12}$,
I.~Allekotte$^{1}$,
A.~Almela$^{8,11}$,
J.~Alvarez Castillo$^{63}$,
J.~Alvarez-Mu\~niz$^{74}$,
G.A.~Anastasi$^{42,43}$
% ...

%% file: latex_authorlist_institutions.tex

\begin{description}[labelsep=0.2em,align=right,labelwidth=0.7em,labelindent=0em,leftmargin=2em,noitemsep]
\item[$^{1}$] Centro At\'omico Bariloche and Instituto Balseiro (CNEA-UNCuyo-CONICET), San Carlos de Bariloche, Argentina
\item[$^{2}$] Centro de Investigaciones en L\'aseres y Aplicaciones, CITEDEF and CONICET, Villa Martelli, Argentina
\item[$^{3}$] Departamento de F\'\i{}sica and Departamento de Ciencias de la Atm\'osfera y los Oc\'eanos, FCEyN, Universidad de Buenos Aires and CONICET, Buenos Aires, Argentina
\item[$^{4}$] IFLP, Universidad Nacional de La Plata and CONICET, La Plata, Argentina
\item[$^{5}$] Instituto de Astronom\'\i{}a y F\'\i{}sica del Espacio (IAFE, CONICET-UBA), Buenos Aires, Argentina
\item[$^{6}$] Instituto de F\'\i{}sica de Rosario (IFIR) -- CONICET/U.N.R.\ and Facultad de Ciencias Bioqu\'\i{}micas y Farmac\'euticas U.N.R., Rosario, Argentina
\item[$^{7}$] Instituto de Tecnolog\'\i{}as en Detecci\'on y Astropart\'\i{}culas (CNEA, CONICET, UNSAM), and Universidad Tecnol\'ogica Nacional -- Facultad Regional Mendoza (CONICET/CNEA), Mendoza, Argentina
\item[$^{8}$] Instituto de Tecnolog\'\i{}as en Detecci\'on y Astropart\'\i{}culas (CNEA, CONICET, UNSAM), Buenos Aires, Argentina
\item[$^{9}$] Observatorio Pierre Auger, Malarg\"ue, Argentina
\item[$^{10}$] Observatorio Pierre Auger and Comisi\'on Nacional de Energ\'\i{}a At\'omica, Malarg\"ue, Argentina
\item[$^{11}$] Universidad Tecnol\'ogica Nacional -- Facultad Regional Buenos Aires, Buenos Aires, Argentina
\item[$^{12}$] University of Adelaide, Adelaide, S.A., Australia
\item[$^{13}$] Universit\'e Libre de Bruxelles (ULB), Brussels, Belgium
\item[$^{14}$] Vrije Universiteit Brussels, Brussels, Belgium
\item[$^{15}$] Centro Brasileiro de Pesquisas Fisicas, Rio de Janeiro, RJ, Brazil
\item[$^{16}$] Centro Federal de Educa\c{c}\~ao Tecnol\'ogica Celso Suckow da Fonseca, Nova Friburgo, Brazil
\item[$^{17}$] Universidade de S\~ao Paulo, Escola de Engenharia de Lorena, Lorena, SP, Brazil
\item[$^{18}$] Universidade de S\~ao Paulo, Instituto de F\'\i{}sica de S\~ao Carlos, S\~ao Carlos, SP, Brazil
\item[$^{19}$] Universidade de S\~ao Paulo, Instituto de F\'\i{}sica, S\~ao Paulo, SP, Brazil
\item[$^{20}$] Universidade Estadual de Campinas, IFGW, Campinas, SP, Brazil
\item[$^{21}$] Universidade Estadual de Feira de Santana, Feira de Santana, Brazil
\item[$^{22}$] Universidade Federal do ABC, Santo Andr\'e, SP, Brazil
\item[$^{23}$] Universidade Federal do Paran\'a, Setor Palotina, Palotina, Brazil
\item[$^{24}$] Universidade Federal do Rio de Janeiro, Instituto de F\'\i{}sica, Rio de Janeiro, RJ, Brazil
\item[$^{25}$] Universidade Federal do Rio de Janeiro (UFRJ), Observat\'orio do Valongo, Rio de Janeiro, RJ, Brazil
\item[$^{26}$] Universidade Federal Fluminense, EEIMVR, Volta Redonda, RJ, Brazil
\item[$^{27}$] Universidad de Medell\'\i{}n, Medell\'\i{}n, Colombia
\item[$^{28}$] Universidad Industrial de Santander, Bucaramanga, Colombia
\item[$^{29}$] Charles University, Faculty of Mathematics and Physics, Institute of Particle and Nuclear Physics, Prague, Czech Republic
\item[$^{30}$] Institute of Physics of the Czech Academy of Sciences, Prague, Czech Republic
\item[$^{31}$] Palacky University, RCPTM, Olomouc, Czech Republic
\item[$^{32}$] Institut de Physique Nucl\'eaire d'Orsay (IPNO), Universit\'e Paris-Sud, Univ.\ Paris/Saclay, CNRS-IN2P3, Orsay, France
\item[$^{33}$] Laboratoire de Physique Nucl\'eaire et de Hautes Energies (LPNHE), Universit\'es Paris 6 et Paris 7, CNRS-IN2P3, Paris, France
\item[$^{34}$] Univ.\ Grenoble Alpes, CNRS, Grenoble Institute of Engineering Univ.\ Grenoble Alpes, LPSC-IN2P3, 38000 Grenoble, France, France
\item[$^{35}$] Bergische Universit\"at Wuppertal, Department of Physics, Wuppertal, Germany
\item[$^{36}$] Karlsruhe Institute of Technology, Institute for Experimental Particle Physics (ETP), Karlsruhe, Germany
\item[$^{37}$] Karlsruhe Institute of Technology, Institut f\"ur Kernphysik, Karlsruhe, Germany
\item[$^{38}$] Karlsruhe Institute of Technology, Institut f\"ur Prozessdatenverarbeitung und Elektronik, Karlsruhe, Germany
\item[$^{39}$] RWTH Aachen University, III.\ Physikalisches Institut A, Aachen, Germany
\item[$^{40}$] Universit\"at Hamburg, II.\ Institut f\"ur Theoretische Physik, Hamburg, Germany
\item[$^{41}$] Universit\"at Siegen, Fachbereich 7 Physik -- Experimentelle Teilchenphysik, Siegen, Germany
\item[$^{42}$] Gran Sasso Science Institute, L'Aquila, Italy
\item[$^{43}$] INFN Laboratori Nazionali del Gran Sasso, Assergi (L'Aquila), Italy
\item[$^{44}$] INFN, Sezione di Catania, Catania, Italy
\item[$^{45}$] INFN, Sezione di Lecce, Lecce, Italy
\item[$^{46}$] INFN, Sezione di Milano, Milano, Italy
\item[$^{47}$] INFN, Sezione di Napoli, Napoli, Italy
\item[$^{48}$] INFN, Sezione di Roma ``Tor Vergata'', Roma, Italy
\item[$^{49}$] INFN, Sezione di Torino, Torino, Italy
\item[$^{50}$] Osservatorio Astrofisico di Torino (INAF), Torino, Italy
\item[$^{51}$] Politecnico di Milano, Dipartimento di Scienze e Tecnologie Aerospaziali , Milano, Italy
\item[$^{52}$] Universit\`a del Salento, Dipartimento di Matematica e Fisica ``E.\ De Giorgi'', Lecce, Italy
\item[$^{53}$] Universit\`a dell'Aquila, Dipartimento di Scienze Fisiche e Chimiche, L'Aquila, Italy
\item[$^{54}$] Universit\`a di Catania, Dipartimento di Fisica e Astronomia, Catania, Italy
\item[$^{55}$] Universit\`a di Milano, Dipartimento di Fisica, Milano, Italy
\item[$^{56}$] Universit\`a di Napoli ``Federico II'', Dipartimento di Fisica ``Ettore Pancini'', Napoli, Italy
\item[$^{57}$] Universit\`a di Roma ``Tor Vergata'', Dipartimento di Fisica, Roma, Italy
\item[$^{58}$] Universit\`a Torino, Dipartimento di Fisica, Torino, Italy
\item[$^{59}$] Benem\'erita Universidad Aut\'onoma de Puebla, Puebla, M\'exico
\item[$^{60}$] Centro de Investigaci\'on y de Estudios Avanzados del IPN (CINVESTAV), M\'exico, D.F., M\'exico
\item[$^{61}$] Unidad Profesional Interdisciplinaria en Ingenier\'\i{}a y Tecnolog\'\i{}as Avanzadas del Instituto Polit\'ecnico Nacional (UPIITA-IPN), M\'exico, D.F., M\'exico
\item[$^{62}$] Universidad Aut\'onoma de Chiapas, Tuxtla Guti\'errez, Chiapas, M\'exico
\item[$^{63}$] Universidad Nacional Aut\'onoma de M\'exico, M\'exico, D.F., M\'exico
\item[$^{64}$] Institute of Nuclear Physics PAN, Krakow, Poland
\item[$^{65}$] University of \L{}\'od\'z, Faculty of Astrophysics, \L{}\'od\'z, Poland
\item[$^{66}$] University of \L{}\'od\'z, Faculty of High-Energy Astrophysics,\L{}\'od\'z, Poland
\item[$^{67}$] Laborat\'orio de Instrumenta\c{c}\~ao e F\'\i{}sica Experimental de Part\'\i{}culas -- LIP and Instituto Superior T\'ecnico -- IST, Universidade de Lisboa -- UL, Lisboa, Portugal
\item[$^{68}$] ``Horia Hulubei'' National Institute for Physics and Nuclear Engineering, Bucharest-Magurele, Romania
\item[$^{69}$] Institute of Space Science, Bucharest-Magurele, Romania
\item[$^{70}$] University Politehnica of Bucharest, Bucharest, Romania
\item[$^{71}$] Center for Astrophysics and Cosmology (CAC), University of Nova Gorica, Nova Gorica, Slovenia
\item[$^{72}$] Experimental Particle Physics Department, J.\ Stefan Institute, Ljubljana, Slovenia
\item[$^{73}$] Universidad de Granada and C.A.F.P.E., Granada, Spain
\item[$^{74}$] Instituto Galego de F\'\i{}sica de Altas Enerx\'\i{}as (I.G.F.A.E.), Universidad de Santiago de Compostela, Santiago de Compostela, Spain
\item[$^{75}$] IMAPP, Radboud University Nijmegen, Nijmegen, The Netherlands
\item[$^{76}$] KVI -- Center for Advanced Radiation Technology, University of Groningen, Groningen, The Netherlands
\item[$^{77}$] Nationaal Instituut voor Kernfysica en Hoge Energie Fysica (NIKHEF), Science Park, Amsterdam, The Netherlands
\item[$^{78}$] Stichting Astronomisch Onderzoek in Nederland (ASTRON), Dwingeloo, The Netherlands
\item[$^{79}$] Universiteit van Amsterdam, Faculty of Science, Amsterdam, The Netherlands
\item[$^{80}$] Case Western Reserve University, Cleveland, OH, USA
\item[$^{81}$] Colorado School of Mines, Golden, CO, USA
\item[$^{82}$] Department of Physics and Astronomy, Lehman College, City University of New York, Bronx, NY, USA
\item[$^{83}$] Louisiana State University, Baton Rouge, LA, USA
\item[$^{84}$] Michigan Technological University, Houghton, MI, USA
\item[$^{85}$] New York University, New York, NY, USA
\item[$^{86}$] Pennsylvania State University, University Park, PA, USA
\item[$^{87}$] University of Chicago, Enrico Fermi Institute, Chicago, IL, USA
\item[$^{88}$] University of Delaware, Bartol Research Institute, Department of Physics and Astronomy, Newark, USA
\item[$^{89}$] University of Nebraska, Lincoln, NE, USA
\item[] -----
\item[$^{a}$] School of Physics and Astronomy, University of Leeds, Leeds, United Kingdom
\item[$^{b}$] Max-Planck-Institut f\"ur Radioastronomie, Bonn, Germany
\item[$^{c}$] Fermi National Accelerator Laboratory, USA
\item[$^{d}$] also at Universidade Federal de Alfenas, Po\c{c}os de Caldas, Brazil
\item[$^{e}$] Colorado State University, Fort Collins, CO, USA
\item[$^{f}$] now at the Hakubi Center for Advanced Research and Graduate School of Science at Kyoto University
\item[$^{g}$] also at Karlsruhe Institute of Technology, Karlsruhe, Germany
\item[$^{h}$] also at University of Bucharest, Physics Department, Bucharest, Romania
\item[$^{i}$] now at European Southern Observatory, Garching bei M\"unchen, Germany
\end{description}